\input mtexsis
\input epsf
\paper
\singlespaced
\widenspacing
\twelvepoint
\Eurostyletrue
\thicksize=0pt
\sectionminspace=0.1\vsize

\def\nc{{N}}
\def\yo1{{F_\pi^2}}
\def\llra{{\relbar\joinrel\longrightarrow}}
\def\mapright#1{{\smash{\mathop{\llra}\limits_{#1}}}}

\def\sss{\scriptscriptstyle}
\def\ssty{\scriptstyle}


\referencelist
\reference{thooft} G.~'t Hooft,
\journal Nucl. Phys. B;72,461 (1974)
\endreference
\reference{witten} E.~Witten,
\journal Nucl. Phys. B;156,269 (1979)
\endreference
\reference{gervais}  J.L.~Gervais and B.~Sakita,
\journal Phys. Rev. Lett.;52,87 (1984);\hfill\break
\journal Phys. Rev. D;30,1795 (1984)
\endreference
\reference{*gervaisa}  R.~Dashen and A.V.~Manohar,
\journal Phys. Lett. B;315,425 (1993), {{\tt hep-ph/9307241}}; 
{\it ibid}, 447, {{\tt hep-ph/9307242}} 
\endreference
\reference{sfsr}  S.~Weinberg, \journal Phys. Rev. Lett.;18,507 (1967)
\endreference
\reference{ope}  C.~Bernard, A.~Duncan, J.~LoSecco and S.~Weinberg, 
\journal Phys. Rev. D;12,792 (1975)
\endreference
\reference{cole} S.~Coleman and E.~Witten,
\journal Phys. Rev. Lett.;45,100 (1980)
\endreference
\reference{derafael} E.~de Rafael and M.~Knecht, \journal Phys. Lett. B;424,335 (1998), {{\tt hep-ph/9712457}}
\endreference
\reference{higgs} See, for instance, {{\it The Standard Model Higgs Boson}}, edited by M.B.~Einhorn, (North-Holland, Amsterdam, 1991)
\endreference 
\reference{algebraic}  S.~Weinberg, \journal Phys. Rev.;177,2604 (1969)
\endreference
\reference{mended}  S.~Weinberg, \journal Phys. Rev. Lett.;65,1177 (1990)
\endreference
\reference{alg} B.~Zumino, in {\it Theory and Phenomenology in Particle
Physics}, 1968 International School of Physics `Ettore Majorana'
(Academic Press, New York, 1969) p.42
\endreference
\reference{*alga} S.~Weinberg, in {\it Lectures on Elementary Particles and
Quantum Field Theory}, edited by Stanley Deser {\it et al}, 
(MIT Press, Cambridge, MA, 1970) p.285
\endreference 
\reference{*algb}  R.~de Alfaro, S.~Fubini, G.~Furlan, and G.~Rossetti, 
            {\it Currents in Hadron Physics}, (North-Holland, Amsterdam, 1973)
\endreference
\reference{mustaki}  D.~Mustaki, {{\tt hep-ph/9410313}}
\endreference
\reference{ecker} G.~Ecker {\it et al},
\journal Phys. Lett. B;223,425 (1989)
\endreference
\reference{pich} F.~Guerrero and A.~Pich,
\journal Phys. Lett. B;412,382 (1997), {{\tt hep-ph/9707347}}
\endreference
\reference{pipi} {{\it Pion-pion Interactions in Particle Physics}},
by B.R.~Martin, D.~Morgan and G.~Shaw, (Academic Press, London, 1976)
\endreference
\reference{gass}  J.~Gasser and H.~Leutwyler, 
\journal Ann. Phys.;158,142 (1984)
\endreference
\reference{froi} M.~Froissart,
\journal Phys. Rev.;123,1053 (1961)
\endreference
\reference{suss}  A.~Casher and L.~Susskind,
\journal Phys. Lett. B;44,171 (1973);\hfill\break
\journal Phys. Rev. D;9,436 (1973)
\endreference
\reference{*sussa} L.~Susskind and M.~Burkardt, {{\tt hep-ph/9410313}}
\endreference
\reference{beanefirst}  S.R.~Beane, in {\it Proceedings of the Fourth Workshop on Quantum Chromodynamics at the AUP}, (World Scientific), to appear, {{\tt hep-ph/9810298}}
\endreference
\reference{beane}  S.R.~Beane, {\it Low-energy Constants from High-energy Theorems}
\endreference
\reference{strong}  S.~Weinberg, {\it Strong Interactions at Low Energies}, {{\tt hep-ph/9412326}}
\endreference
\reference{beane2}  S.R.~Beane, \journal Phys. Rev. D;59,031901(1999), {{\tt hep-ph/9809328}}
\endreference
\reference{luty} M.A.~Luty and R.~Sundrum, \journal Phys. Rev. D;52,1627 (1995), {{\tt hep-ph/9502259}}
\endreference
\endreferencelist

\titlepage
\obeylines
\hskip4.8in{DOE/ER/40762-170}
\hskip4.8in{U.ofMd.PP\#99-054}
\hskip4.8in{NT@UW-99-53}\unobeylines
\vskip0.5in
\title
High Energy Theorems at Large-$\nc$
\endtitle
\author
Silas R.~Beane

Department of Physics, University of Maryland
College Park, MD 20742-4111
\vskip0.05in
\center{{\it and}}\endcenter
\vskip0.05in
Department of Physics, University of Washington
Seattle, WA 98195-1560
\vskip0.1in
\center{{\it sbeane@phys.washington.edu}}\endcenter
\endauthor

\abstract
\singlespaced
\widenspacing

Sum rules for products of two, three and four QCD currents are derived
using chiral symmetry at infinite momentum in the large-$\nc$
limit. These exact relations among meson decay constants, axialvector
couplings and masses determine the asymptotic behavior of an infinite
number of QCD correlators.  The familiar spectral function sum rules
for products of two QCD currents are among the relations derived. With
this precise knowledge of asymptotic behavior, an infinite number of
large-$\nc$ QCD correlators can be constructed using dispersion
relations.  A detailed derivation is given of the exact large-$\nc$
pion vector form factor and forward pion-pion scattering amplitudes.

\endabstract 
\vskip0.5in
\center{PACS: 11.30.Rd; 12.38.Aw; 11.55.Jy; 11.30.Er} 
\endcenter
\endtitlepage
\vfill\eject                                     
\superrefsfalse
\singlespaced
\widenspacing

\vskip0.1in
\noindent {\twelvepoint{\bf 1.\quad Introduction}}
\vskip0.1in

In the large-$\nc$ limit all QCD mesonic correlation functions are
sums of tree graphs\ref{thooft}\ref{witten}. The machinery of
dispersion relations, which relates correlators at different energies,
is transparent in this context since all dispersion relations are
saturated by single-particle states. Of course dispersion relations
require knowledge of asymptotic behavior and traditionally this
knowledge has been assumed or taken from phenomenologically inspired
models like Regge pole theory. What is lacking is a fundamental
principle based in QCD which determines the asymptotic behavior of
correlation functions. The dominance of single-particle states at
large-$\nc$ has a related advantage: this limit is particularly suited
to a symmetry point of view since it is easy to place single-particle
states in representations of a symmetry group and extract consequences
for matrix elements. In fact, there is a deep relation between
asymptotic behavior and symmetries.  For instance, requiring
asymptotic behavior consistent with unitarity implies a spin-flavor
algebra which constrains the spectrum of the large-$\nc$
baryons\ref{gervais}. Another classic example is string theory, where
both the spectrum of states and the asymptotic behavior of S-matrix
elements (which are QCD-like) are determined by symmetries on the
string world sheet. In this paper I show that in large-$\nc$ QCD,
chiral symmetry determines the asymptotic behavior of an infinite
number of mesonic correlation functions.  Armed with this asymptotic
information, exact large-$\nc$ correlators can be constructed using
dispersion relations in a facile way.

There are well known QCD predictions of asymptotic behavior for
products of two currents with nontrivial chiral quantum
numbers\ref{sfsr}.  These asymptotic constraints ---known as spectral
function sum rules (SFSR's)--- are naturally derived in the context of
the operator product expansion (OPE)\ref{ope}. The fundamental
ingredient of their derivation is one of the remarkable properties of
the OPE: coefficient functions in the OPE are insensitive to vacuum
properties and therefore transform with respect to the full unbroken
global symmetry group of the underlying theory, regardless of whether
this symmetry is spontaneously broken in the low-energy
theory\ref{ope}. Hence the SFSR's exist because the full global chiral
symmetry group of QCD can be used for classification purposes in the
region of large Euclidean momenta. One might then wonder whether the
SFSR's can be derived directly using chiral symmetry without recourse
to the OPE. In full QCD a direct symmetry approach is hindered by
complicated analyticities arising from various continuum contributions
to the two-point functions. On the other hand, in the large-$\nc$
limit, the two-point functions are determined by the exchange of an
infinite number of single-particle states.  I will show that the
SFSR's at large-$\nc$ are {\it purely} consequences of chiral symmetry
and are easily derived directly from the chiral representation theory
and/or the chiral algebra without recourse to the OPE. Relevant in
this respect is a special class of Lorentz frames in which a component
of momentum is taken to be very large compared to typical hadronic
scales (e.g. the infinite momentum frame).

The purpose of this work is not to find an amusing way of reproducing
the SFSR's. Rather it is to see if the direct symmetry approach leads
to other sum rules which have been overlooked or simply obscured by
the complexities of the OPE.  I will derive an infinite number of
(new) sum rules in the large-$\nc$ limit. These sum rules rely on
precisely the same theoretical assumptions as the SFSR's and in fact
depend on the SFSR's for consistency.  Just as the SFSR's determine
the asymptotic behavior of two-point functions which in turn can be
constructed using dispersion relations, all of the new chiral sum
rules determine the asymptotic behavior of specific three- or
four-point functions in large-$\nc$ QCD. Dispersion relations can then
be used to construct the exact correlators.  I will consider several
examples in detail.

In section 2 I review some basic relevant facts about large-$\nc$ QCD.
The technology of the operator product expansion (OPE) is then used to
review the derivation of the spectral function sum rules. In section 3
I derive the SFSR's and an infinite number of new sum rules directly
using chiral symmetry and its representation theory in the infinite
momentum frame. In section 4 I identify some of these sum rules with
asymptotic constraints and explicitly construct exact large-$\nc$
correlators using dispersion relations. I conclude in section 5. In an
appendix I derive a set of sum rules directly from the chiral algebra.

\vskip0.1in
\noindent {\twelvepoint{\bf 2.\quad Spectral Function Sum Rules at Large-$\nc$}}
\vskip0.1in

This section reviews known material which is essential for what
follows.  In this paper I consider QCD with two massless flavors.
This theory has an $SU(2)_L\times SU(2)_R$ chiral symmetry.  In the
large-$\nc$ limit, assuming confinement, this symmetry breaks
spontaneously to the $SU(2)_V$ isospin subgroup\ref{cole}. (Strictly
speaking, large-$\nc$ QCD has a $U(2)_L\times U(2)_R$ chiral symmetry.
In this paper the additional Goldstone mode and its associated $U(1)$
charge are ignored.) In the large-$\nc$ limit the mesons have the most
general quantum numbers of the quark bilinear ${\bar {\rm Q}}\Gamma
{\rm Q}$ where $\Gamma$ is some arbitrary spin structure.  This means
that all mesons have zero or unit isospin. The corresponding
chiral transformation properties of the mesons will be discussed
below. The $SU(2)\times SU(2)$ algebra is expressed via the
commutation relations

\offparens
$$
[{Q^{5}_{a}},{Q^{5}_{b}}]=i{\epsilon_{abc}}{T_c} 
\qquad
[{T^{\;\;}_a},{Q^{5}_{b}}]=
i{\epsilon_{abc}}{Q^{5}_{c}}
\qquad
[{T^{\;\;}_a},{T^{\;\;}_b}]=i{\epsilon_{abc}}{T^{\;\;}_c},
\EQN chiralalg
$$\autoparens where ${T^a}$ are $SU(2)_V$ generators normalized so
that ${T^a}={\tau_a}/2$ and ${Q^{5}_{a}}$ are the $SU(2)_A$ generators
that are broken by the vacuum, at rest. Note that $V=L+R$ and $A=L-R$.
The conserved QCD currents associated with these charges are:

\offparens
$$\eqalign{ 
&{{\rm A}_{\mu}^a}=
{\bar {\rm Q}}{\gamma_\mu}{\gamma_5}{T^a}{\rm Q}\cr
&{{\rm V}_{\mu}^a}=
{\bar {\rm Q}}{\gamma_\mu}{T^a}{\rm Q}.\cr}
\EQN aandvdefinqcd$$ 
These currents satisfy the commutation relations

\offparens
$$
[{Q_{5}^{a}},{{\rm V}_{\mu}^b}]=i{\epsilon^{abc}}{{\rm A}_{\mu}^c}
\qquad
[{Q_{5}^{a}},{{\rm A}_{\mu}^b}]=i{\epsilon^{abc}}{{\rm V}_{\mu}^c}.
\EQN currentalg1
$$\autoparens Therefore, ${{\rm V}_{\mu}^a}$ and ${{\rm A}_{\mu}^a}$
form a complete (six dimensional) chiral multiplet; they transform as
$(\bf{3},\bf{1})\oplus (\bf{1},\bf{3})$ with respect to $SU({2})\times
SU({2})$.

In the low-energy theory, the axialvector current ${{\rm A}_{\mu}^a}$
has a nonvanishing matrix element between the vacuum and the Goldstone
boson (pion) states:

$$
\bra{0}{{\rm A}_{\mu}^a}\ket{\pi^b}=
{\delta^{ab}}{F_{\pi}}{p_\mu}.
\EQN vacgold$$
Both conserved currents have nonvanishing matrix elements between the
vacuum and vector meson states:

\offparens
$$\eqalign{ 
&\bra{0}{{\rm A}_{\mu}^a}\ket{A^{b}}^{\sss (\lambda )}=
{\delta^{ab}}{F_{\sss A}}{M_{\sss A}}{\epsilon_\mu^{\sss (\lambda )}}\cr
&\bra{0}{{\rm V}_{\mu}^a}\ket{V^{b}}^{\sss (\lambda )}=
{\delta^{ab}}{F_{\sss V}}{M_{\sss V}}{\epsilon_\mu^{\sss (\lambda )}}\cr}
\EQN aandvdef$$ 
where $\epsilon_\mu^{\sss (\lambda )}$ is the vector meson
polarization vector and $\lambda$ is a helicity label. 

Consider the time-ordered product of currents:

$$
{\Pi_{\sss LR}^{\mu\nu}}(q){\delta_{ab}}=
4i\int {d^4x}{e^{iqx}}
\bra{0}T{L_{a}^\mu}(x){R_{b}^\nu}(0)\ket{0},
\EQN correl1$$ 
where the $SU(2)_{L,R}$ QCD currents $L$ and $R$ transform as
$(\bf{3},\bf{1})$ and $(\bf{1},\bf{3})$, respectively, with respect to
$SU(2)\times SU(2)$. It follows that ${\Pi_{\sss LR}}$ transforms as
$(\bf{3},\bf{3})$. Lorentz invariance and current conservation allow
the decomposition

$${\Pi_{\sss LR}^{\mu\nu}}(q)=
({q^\mu}{q^\nu}-{g^{\mu\nu}}{q^2}){\Pi_{\sss LR}}({Q^2}),
\EQN correl2$$ 
where ${Q^2}=-{q^2}$.  Because the function ${\Pi_{\sss LR}}({Q^2})$
carries nontrivial chiral quantum numbers, it vanishes to all orders
in QCD perturbation theory. Hence, asymptotic freedom implies
${\Pi_{\sss LR}}({Q^2})\rightarrow 0$ as ${Q^2}\rightarrow\infty$ and
${\Pi_{\sss LR}}$ satisfies the unsubtracted dispersion relation:

$$
{\Pi_{\sss LR}}({Q^2})
={1\over{\pi}}
\int{d{t}\;{Im\; {{\Pi_{\sss LR}}({t})}}\over{{t}+{Q^2}}}.
\EQN kallenl$$ 
In the OPE we have the expansion,

$$
{\Pi_{\sss LR}}({Q^2})
\mapright{{Q^2}\rightarrow\infty}\, \,
{\sum_{n=1}^{\infty}} 
{{\vev{{\cal O}}^{\sss d=2n}_{\sss \bf{(3,3)}}}\over{Q^{2n}}}
\EQN correl3$$ 
where the coefficients ${{\vev{{\cal O}}^{\sss d=2n}_{\sss (3,3)}}}$,
of mass dimension $d=2n$ transform as $(\bf{3},\bf{3})$ with respect
to $SU(2)\times SU(2)$. {\it This is the case regardless of whether
$SU(2)\times SU(2)$ is spontaneously broken in the low-energy theory.}
The coefficient functions always transform with respect to the full
unbroken symmetry group of the underlying theory\ref{ope}.

In the large-$\nc$ limit ${\Pi_{\sss LR}}$ is given by a sum of
single-particle states\ref{derafael}. Using \Eq{vacgold}-\Eq{correl2}
one obtains

$$
-{1\over\pi}{Im\; {{\Pi_{\sss LR}}({t})}}=
\yo1\delta ({t})+
{\sum _{\sss A}}{F_{\sss A}^2}\delta (t-{M_{\sss A}^2})-
{\sum _{\sss V}}{F_{\sss V}^2}\delta (t-{M_{\sss V}^2}).
\EQN kallenlargensat$$ 
Evaluating the dispersion relation then gives

$$
-{Q^2}\;{\Pi_{\sss LR}}({Q^2})=\yo1 +
{\sum _{\sss A}}
{{{Q^2}{F_{\sss A}^2}}\over{{Q^2}+{M_{\sss A}^2}}}-
{\sum _{\sss V}}
{{{Q^2}{F_{\sss V}^2}}\over{{Q^2}+{M_{\sss V}^2}}},
\EQN sol$$ 
which can be expanded in inverse powers of ${Q^2}$:

\offparens
$$
{\Pi_{\sss LR}}({Q^2})
\mapright{{Q^2}\rightarrow\infty}\, 
\Bigl\{{\sum _{\sss V}}{F_{\sss V}^2}-
{\sum _{\sss A}}{F_{\sss A}^2}-\yo1\Bigr\}{1\over{Q^2}}
+{\sum_{\sss m=1}^\infty}
\Bigl\{{\sum_{\sss V}}{F_{\sss V}^2}{M_{\sss V}^{2m}}-
{\sum_{\sss A}}{F_{\sss A}^2}{M_{\sss A}^{2m}}
\Bigr\}{{{(-1)^m}}\over{Q^{2m+2}}}.
\EQN kallenasymbehfulfu$$
The coefficients in this expansion can be matched to the OPE, giving

\offparens
$$\EQNalign{ 
&{\vev{{\cal O}}^{\sss d=2}_{\sss (3,3)}}= 
{\sum _{\sss V}}{F_{\sss V}^2}-
{\sum _{\sss A}}{F_{\sss A}^2}-\yo1
\EQN presfsrex;a\cr
&{\vev{{\cal O}}^{\sss d=2n}_{\sss (3,3)}}= 
{{(-1)^{n-1}}}\Bigl\{
{\sum_{\sss V}}{F_{\sss V}^2}{M_{\sss V}^{2n-2}}-
{\sum_{\sss A}}{F_{\sss A}^2}{M_{\sss A}^{2n-2}}\Bigr\}\qquad n>1.
\EQN presfsrex;b\cr}
$$
There are no QCD gauge invariant local operators with $d=2$. The only
$d=4$ operator is $F^2$ where $F$ is the gluon field strength
tensor. However, this operator is a chiral singlet. Hence the first
two terms in the OPE must vanish and we have the two spectral function
sum rules:

\offparens
$$\EQNalign{ 
&{\vev{{\cal O}}^{\sss d=2}_{\sss (3,3)}}=0 
\qquad\Longrightarrow\qquad
{\sum _{\sss V}}{F_{\sss V}^2}=
{\sum _{\sss A}}{F_{\sss A}^2}+\yo1
\EQN sfsrex;a\cr
&{\vev{{\cal O}}^{\sss d=4}_{\sss (3,3)}}=0 
\qquad\Longrightarrow\qquad
{\sum_{\sss V}}{F_{\sss V}^2}{M_{\sss V}^{2}}=
{\sum_{\sss A}}{F_{\sss A}^2}{M_{\sss A}^{2}},
\EQN sfsrex;b\cr}
$$
which are referred to in the text as {\it SFSR1} and {\it SFSR2},
respectively.  These sum rules are illustrated diagrammatically in
\Fig{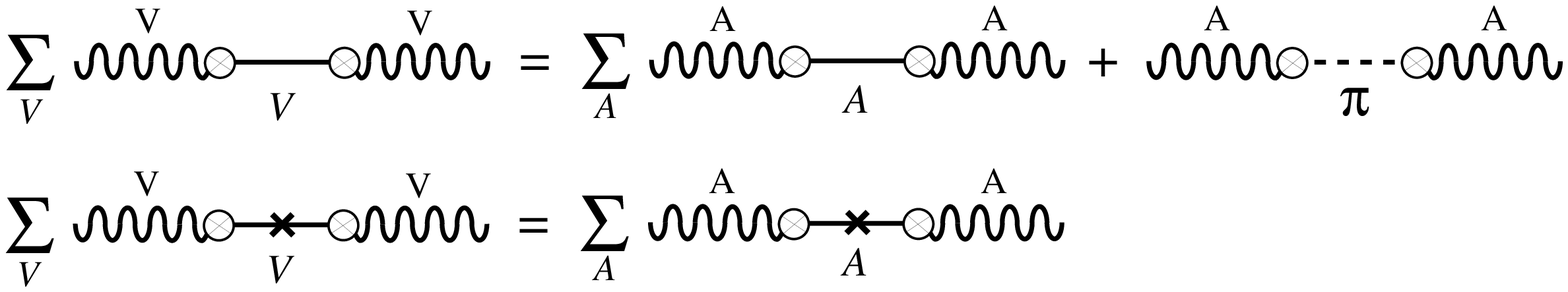}.  Of course the sums span an infinite number of
states in the large-$\nc$ limit$^1$\vfootnote1{Some interesting
consequences of these sum rules when truncated to a small number of
states are discussed in Ref.~\ref{derafael}.}.  The exact large-$\nc$
two-point function may then be written as

$$
{\Pi_{\sss LR}}({q^2})=
{\sum _{\sss V}}
{{{F_{\sss V}^2}{M_{\sss V}^4}}\over{{q^4}({M_{\sss V}^2}-{q^2})}}-
{\sum _{\sss A}}
{{{F_{\sss A}^2}{M_{\sss A}^4}}\over{{q^4}({M_{\sss A}^2}-{q^2})}},
\EQN fixedsol$$ 
which manifests the correct asymptotic behavior.

Since narrow single-particle states are all that survive in the
large-$\nc$ limit, one may wonder whether it is possible to obtain the
SFSR's directly using $SU(2)\times SU(2)$ symmetry, with no reference
to the OPE.  One might expect that the pion together with the infinite
number of vector and axialvector states in some sense fill out a
representation of the full unbroken $SU(2)\times SU(2)$ group. We will
see that this is the case.  I reiterate that the main motivation of
the symmetry approach is to find sum rules for other $n$-point
functions in a simple way. For instance, consider adding an external
pion line to one of the diagrams in \Fig{twocurrents.eps} so that a
vector current turns into an axial current (see
\Fig{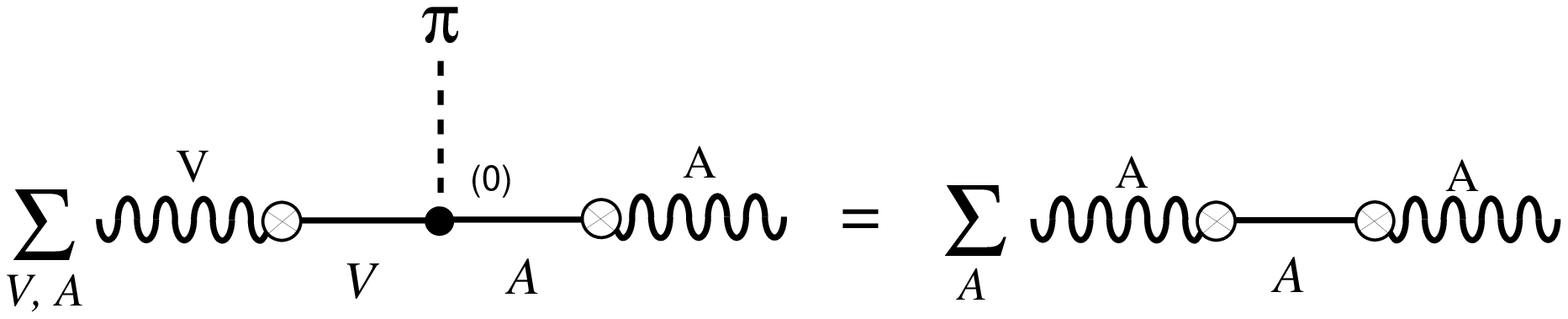}). Intuitively is would seem that there should
be a sum rule analogous to the SFSR's for such a graph. This is in
fact the case, as we will see.

\figure{twocurrents.eps}
\epsfxsize 5in
\centerline{\epsfbox{twocurrents.eps}}
\caption{Spectral function sum rules for 
products of two QCD currents. The top diagram is {\it SFSR1} and the
bottom diagram is {\it SFSR2}.  The cross represents a mass-squared
insertion.}
\endcaption
\endfigure

\vskip0.1in
\noindent {\twelvepoint{\bf 3.\quad Chiral Symmetry at Infinite Momentum}}
\vskip0.1in

\vskip0.1in
\noindent {\twelvepoint{\it 3.1\quad The Infinite Momentum Frame}}

The fact that OPE coefficients do not feel vacuum properties enables
the derivation of the SFSR's by making use of the full unbroken chiral
symmetry group\ref{ope}.  In order to obtain the SFSR's using symmetry
arguments it is necessary to work directly with the matrix
elements of the currents between the states and the vacuum, rather
than with the correlation function ${\Pi_{\sss LR}}$. But clearly the
manner in which the meson decay constants are defined through the
matrix elements of \Eq{vacgold} and \Eq{aandvdef} implies an asymmetry
between states of different spin.  The symmetric appearance of the
decay constants in the SFSR's (for instance, {\it SFSR1} is invariant
with respect to ${F_{\sss A}}\leftrightarrow{F_\pi}$ for each $A$) is
a consequence of taking $Q^2 \rightarrow \infty$ in the OPE. In
studying matrix elements it is therefore convenient to work in a
Lorentz frame in which the pions and the vector mesons appear the
same. This is easy to achieve. At rest we define the vector
meson polarization vectors:

$$
\epsilon_\mu^{\sss (+)}=(0,1,0,0)\qquad
\epsilon_\mu^{\sss (-)}=(0,0,1,0)\qquad
\epsilon_\mu^{\sss (0)}=(0,0,0,1).
\EQN polvecsatrest$$
Consider boosting all particles along the $3$-axis (or the observer
along the negative $3$-axis) to ${p_\mu}=({p_0},0,0,{p_3})$.  We then
have $\epsilon_\mu^{\sss (\pm )}$ unchanged and

$$
\epsilon_\mu^{\sss (0)}=({{p_3}\over M},0,0,{{p_0}\over M}).
\EQN polvecsatmoving$$
Now as we take ${p_3}\rightarrow\infty$,

$$
\epsilon_\mu^{\sss (0)}={{p_\mu}\over M} +O({M\over{p_3}})
\EQN polvecsatmoving$$
and we have

\offparens
$$\eqalign{ 
&\bra{0}{{\rm A}_{\mu}^a}\ket{\pi^b}=
{\delta^{ab}}{F_{\pi}}{p_\mu}\cr
&\bra{0}{{\rm A}_{\mu}^a}\ket{A^{b}}^{\sss (0)}=
{\delta^{ab}}{F_{\sss A}}{p_\mu}\cr 
&\bra{0}{{\rm V}_{\mu}^a}\ket{V^{b}}^{\sss (0)}=
{\delta^{ab}}{F_{\sss V}}{p_\mu}. \cr}
\EQN aandvdefnew$$ 
Therefore, as the momentum is taken large compared to the mass scales
in the problem, the $\lambda =0$ vector mesons act like
Goldstone bosons.  This is familiar from the Goldstone boson
equivalence theorem in electroweak physics\ref{higgs}. The kinematical
conditions described above are known as the infinite momentum
frame. One advantage of this frame is that boost invariance is
preserved along the $3$-direction.

\figure{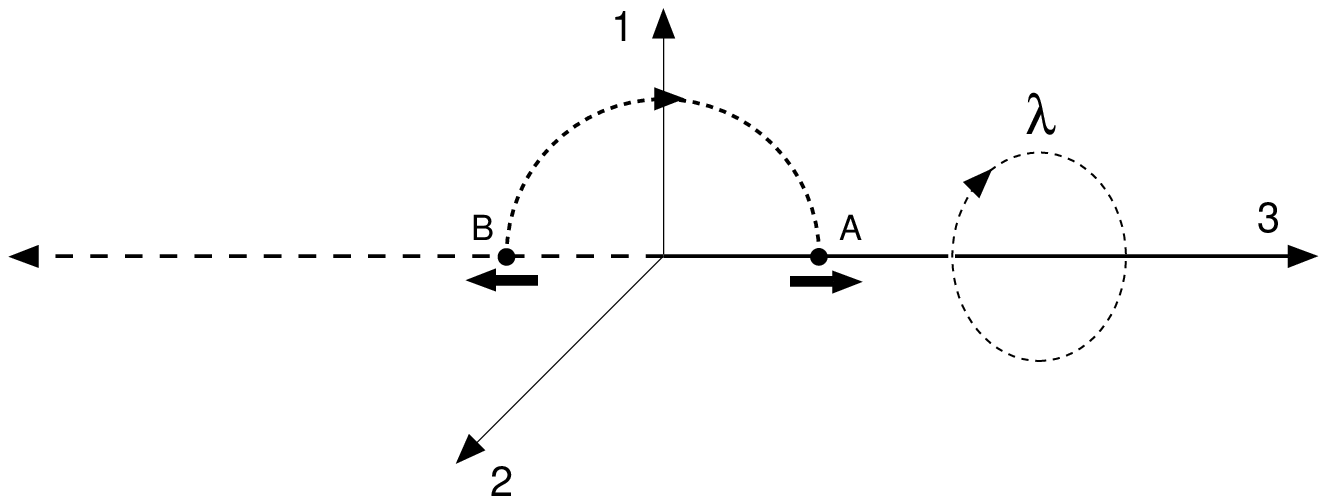}
\epsfxsize 5in
\centerline{\epsfbox{infinitemom.eps}}
\caption{Collinear kinematics of the infinite momentum frame where
${p_3}\rightarrow\infty$.  Conservation of helicity, $\lambda$,
follows from invariance with respect to rotations about the
$3$-axis. Parity takes $\rm A$ to $\rm B$. A rotation by $\pi$ about
the $2$-axis restores the original configuration with the sign of the
spin (helicity) reversed.}
\endcaption
\endfigure

We can use boost invariance to define the amplitudes:

\offparens
$$\eqalign{ 
&\bra{0}{{A_0^a}-{A_3^a}}\ket{\alpha}\equiv
({{p_0}-{p_3}})\bra{0}{{\cal A}^a}\ket{\alpha}\cr
&\bra{0}{{V_0^a}-{V_3^a}}\ket{\alpha}\equiv
({{p_0}-{p_3}})\bra{0}{{\cal V}^a}\ket{\alpha},\cr}
\EQN app4$$ 
where the matrix elements of ${{\cal V}^a}$ and ${{\cal A}^a}$ are
constants and $\alpha$ represents a physical meson state. It then
follows from \Eq{aandvdefnew} and \Eq{app4} that

\offparens
$$\eqalign{ 
&\bra{0}{{\cal A}_a}\ket{\pi_b}=
{\delta_{ab}}{F_{\pi}}\cr
&\bra{0}{{\cal A}_a}\ket{A_b}^{\sss (0)}=
{\delta_{ab}}{F_{\sss A}}\cr
&\bra{0}{{\cal V}_a}\ket{V_b}^{\sss (0)}=
{\delta_{ab}}{F_{\sss V}}.\cr}
\EQN infdecayone$$ 
To reiterate, the matrix elements of the pions and the $\lambda =0$
components of the vector mesons look the same in the infinite momentum
frame. This is ideal when using symmetry arguments to relate matrix
elements.

\vskip0.1in
\noindent {\twelvepoint{\it 3.2\quad Selection Rules at Infinite Momentum}}

We have the following selection rules at infinite 
momentum$^2$\vfootnote2{This formalism is worked out in great detail
in \Ref{algebraic}, \Ref{mended} and \Ref{alg}.}:

\noindent$\bullet$ Invariance with respect to rotations about the
$3$-axis implies helicity conservation. We can look at sectors of
states labelled by helicity. This explains why the $\lambda =0$
components of the vector mesons are on par with the pions at infinite
momentum; all $\lambda =0$ states look like scalars.  We will be
interested in the helicity conserving pion transition matrix elements
between physical meson states $\beta$ and $\alpha$ which are related
to the axial charges in the infinite momentum frame by:

\offparens
$$ {{\cal M}_a}(p'{\lambda '}\beta ,p{\lambda }\alpha )=
{{({{F_\pi}})}^{-1}} ({m_\alpha ^2}-{m_\beta ^2})
\lbrack {^{\sss (\lambda )}\bra{\,\beta\,}}
{Q^{\sss 5}_a}{{\ket{\,\alpha\,}}^{\sss (\lambda )}}\rbrack
{\delta _{{\lambda '}\lambda}}.
\EQN amp $$\autoparens

\vskip0.2in

\noindent$\bullet$ The axial charges annihilate the vacuum in the 
broken phase:

$$
{Q_5^a}\ket{0}=0.
\EQN annih
$$ This means that the chiral algebra is useful for classifying hadron
states$^3$\vfootnote3{An easy way to see this is to express the QCD
axial charges in light-front coordinates. Then it is clear that
${Q_5^a}$ does not carry vacuum quantum numbers\ref{mustaki} and so
annihilates the vacuum.}. We will discuss the manner in which symmetry
breaking effects appear in this frame below.  Since meson states are
quark bilinears in the large-$\nc$ limit and quarks transform as
$(\bf{2},\bf{1})$ and $(\bf{1},\bf{2})$, mesons transform as
combinations of $(\bf{2},\bf{2})$, $(\bf{3},\bf{1})$,
$(\bf{1},\bf{3})$ and $(\bf{1},\bf{1})$ irreducible representations of
$SU({2})\times SU({2})$. Charge conjugation leaves $(\bf{2},\bf{2})$
and $(\bf{1},\bf{1})$ unchanged and interchanges $(\bf{1},\bf{3})$ and
$(\bf{3},\bf{1})$.  Physical meson states have definite charge
conjugation, $C$, and isospin, $I$, and therefore are linear
combinations of the isovectors $\ket{{\bf (2,2)}_a}$, $\{{\ket{{\bf
(1,3)}_a}}- {\ket{{\bf
(3,1)}_a}}\}/{\textstyle{\sqrt{2}}}\equiv{\ket{{\bf V}_a}}$ and
$\{{\ket{{\bf (1,3)}_a}}+ {\ket{{\bf
(3,1)}_a}}\}/{\textstyle{\sqrt{2}}}\equiv{\ket{{\bf A}_a}}$ and the
isoscalars $\ket{{\bf (2,2)}_{\sss 0}}$ and $\ket{\bf{(1,1)}}$.  Roman
subscripts are isospin indices. Only $\ket{{\bf V}_a}$ changes sign
under charge conjugation.  The action of the generators on the states
of definite chirality can be obtained using tensor analysis:

\offparens
$$\EQNalign{ 
&{_i}\bra{{\bf A}_a}{Q^{5}_{b}}\ket{{\bf V}_c}_j=
i{\epsilon_{abc}}{\delta_{ij}}
\qquad {_i}\bra{{\bf (2,2)}_a}{Q^{5}_{b}}\ket{{\bf (2,2)}_{\sss 0}}_j=
i{\delta_{ab}}{\delta_{ij}}
\EQN gentensor;a \cr
&{_i}\bra{{\bf (2,2)}_a}{T_{b}}\ket{{\bf (2,2)}_c}_j=
{_i}\bra{{\bf A}_a}{T_{b}}\ket{{\bf A}_c}_j=
{_i}\bra{{\bf V}_a}{T_{b}}\ket{{\bf V}_c}_j=
i{\epsilon_{abc}}{\delta_{ij}}.
\EQN gentensor;b  \cr}
$$ 
\autoparens

\vskip0.2in

\noindent$\bullet$ Invariance with respect to a combined space inversion and 
rotation through $\pi$ about an axis perpendicular to the $3$-axis implies:

$$
{^{\sss (\lambda )}\bra{\,\alpha\,}}
{Q^{\sss 5}_a}{{\ket{\,\beta\,}}^{\sss (\lambda )}}=
{^{\sss (-\lambda )}\bra{\,\alpha\,}}
{Q^{\sss 5}_a}{{\ket{\,\beta\,}}^{\sss (-\lambda )}}
{P_\alpha}{P_\beta}{{(-1)}^{{J_\beta}-{J_\alpha}+1}}
\EQN normality
$$
where $P$ and $J$ are parity and spin, respectively. For $\lambda =0$ this
implies the selection rule:

$$
{\eta_\beta}=-{\eta_\alpha}
\EQN normality2
$$
else $\bra{\alpha}{Q^{\sss 5}_a}{{\ket{\beta}}}=0$, where $\eta\equiv
{P{{(-1)}^{J}}}$ is normality.  From \Eq{amp} it then follows that
only ($\lambda =0$) states of opposite normality communicate by pion
exchange.

\vskip0.2in

\noindent$\bullet$ Conservation of $G$-parity implies the
selection rule

$$
{G_\beta}=-{G_\alpha}
\EQN gpar
$$ 
else $\bra{\alpha}{Q^{\sss 5}_a}{{\ket{\beta}}}=0$.  The product of
normality and $G$-parity is a symmetry. Hence it follows that
physical meson states fall into representations labelled by $G\eta$.

A geometric picture of several of the selection rules is given in 
\Fig{infinitemom.eps}.

\vskip0.2in

\table{gl}
\tenpoint
\caption{\twelvepoint Members of the pion $SU({2})\times SU({2})$ 
representation at infinite momentum.  The $\lambda=0$ components of
meson states of arbitrary spin with allowed quantum numbers can participate
in this representation.}
\doublespaced
\ruledtable
{}    | $I$  | $G$ | $J$    | $P$ | $C$ | $\eta$ | $G\eta$ \cr
$\pi$ | $1$  | $-$ | $0$ | $-$ | $+$ | $-$  | $+$ \cr
$P$ | $1$  | $-$ | $\it even$ | $-$ | $+$ | $-$  | $+$ \cr
$A$   | $1$  | $-$ | $\it odd$  | $+$ | $+$ | $-$  | $+$ \cr
$V$   | $1$  | $+$ | $\it odd$  | $-$ | $-$ | $+$  | $+$ \cr
$S$   | $0$  | $+$ | $\it even$ | $+$ | $+$ | $+$  | $+$
\endruledtable
\endtable

\vskip0.1in
\noindent {\twelvepoint{\it 3.3\quad The Representation Theory}}

Using the infinite momentum frame selection rules it is
straightforward to identify the {\it most general} pion $SU({2})\times
SU({2})$ representation in the large-$\nc$ limit. Since the pion is a
Lorentz scalar all states in its chiral representation have $\lambda
=0$.  This representation can contain meson states of any spin. Of
course all mesons have $\lambda =0$ components.  Too, the pion has
$G\eta=+1$ as do all states in its representation.  Since the pion has
$I=1$ and $C=+1$ it is a linear combination of $\ket{{\bf (2,2)}_a}$
and $\ket{{\bf A}_a}$ states.  The other physical states which are
linear combinations of $\ket{{\bf (2,2)}_a}$ and $\ket{{\bf A}_a}$
states necessarily have $\eta =-1$ and $G=-1$ but can have $P=-1$,
$J=$even {\it or} $P=+1$, $J=$odd. The former are labelled $P$ and the
latter $A$.  The physical states with $\eta =+1$ and $G=+1$ can have
$P=-1$, $J=$odd {\it or} $P=+1$, $J=$even. Bose statistics requires
that the former have $I=1$ (and thus be linear combinations of
$\ket{{\bf V}_a}$ states with $C=-1$) and the latter have $I=0$ (and
thus be linear combinations of $\ket{{\bf (2,2)}_{\sss 0}}$ and
$\ket{\bf{(1,1)}}$ states with $C=+1$).  These states are labelled $V$
and $S$, respectively.  Allowed states are listed in
\Tbl{gl} together with their quantum numbers.  Thus, the most general
chiral representation involving the pion and these states is:

\offparens 
$$\EQNalign{ 
&\ket{\pi_a}=
{\sum_{\sss i=1}^{\sss m}}{u_{\sss{1i}}}{\ket{{\bf{A}}_a}_i}+
{\sum_{\sss i=m+1}^{\sss n+m}}{u_{\sss{1i}}}{\ket{{\bf{(2,2)}}_a}_i}
\EQN genrep;a\cr
&\ket{A_a}^{\sss (0)}_{\sss 2}=
{\sum_{\sss i=1}^{\sss m}}{u_{\sss{2i}}}{\ket{{\bf{A}}_a}_i}+
{\sum_{\sss i=m+1}^{\sss n+m}}{u_{\sss{2i}}}{\ket{{\bf{(2,2)}}_a}_i} \cr
&\qquad\qquad\vdots     \cr 
&\ket{A_a}^{\sss (0)}_{\sss l}=
{\sum_{\sss i=1}^{\sss m}}{u_{\sss{li}}}{\ket{{\bf{A}}_a}_i}+
{\sum_{\sss i=m+1}^{\sss n+m}}{u_{\sss{li}}}{\ket{{\bf{(2,2)}}_a}_i} 
\EQN genrep;b\cr
&\ket{P_a}^{\sss (0)}_{\sss (l+1)}=
{\sum_{\sss i=1}^{\sss m}}{u_{\sss{(l+1)i}}}{\ket{{\bf{A}}_a}_i}+
{\sum_{\sss i=m+1}^{\sss n+m}}{u_{\sss{(l+1)i}}}{\ket{{\bf{(2,2)}}_a}_i} \cr
&\qquad\qquad\vdots     \cr 
&\ket{P_a}^{\sss (0)}_{\sss n+m}=
{\sum_{\sss i=1}^{\sss m}}{u_{\sss{(n+m)i}}}{\ket{{\bf{A}}_a}_i}+
{\sum_{\sss i=m+1}^{\sss n+m}}{u_{\sss{(n+m)i}}}{\ket{{\bf{(2,2)}}_a}_i} 
\EQN genrep;c\cr
&\ket{{V}_a}^{\sss (0)}_{\sss 1}=
{\sum_{\sss j=1}^{\sss m}}{w_{\sss{1j}}}{\ket{{\bf{V}}_a}_j}\cr
&\qquad\qquad\vdots\cr
&\ket{{V}_a}^{\sss (0)}_{\sss m}=
{\sum_{\sss j=1}^{\sss m}}{w_{\sss{mj}}}{\ket{{\bf{V}}_a}_j}
\EQN genrep;d\cr
&\ket{S}^{\sss (0)}_{\sss 1}=
{\sum_{\sss i=m+1}^{\sss n+m}}{v_{\sss{(m+1)i}}}{\ket{{\bf{(2,2)}}_{\sss 0}}_i}
+{\sum_{\sss i=m+n+1}^{\sss m+n+o}}{v_{\sss{(m+1)i}}}{\ket{{\bf{(1,1)}}}_i}\cr
&\qquad\qquad\vdots\cr
&\ket{S}^{\sss (0)}_{\sss n+o}=
{\sum_{\sss i=m+1}^{\sss m+n}}
{v_{\sss{(m+n+o)i}}}{\ket{{\bf{(2,2)}}_{\sss 0}}_i}
+{\sum_{\sss i=m+n+1}^{\sss m+n+o}}{v_{\sss{(m+n+o)i}}}{\ket{{\bf{(1,1)}}}_i}
\EQN genrep;e\cr}
$$ 
\autoparens
where $\hat u$, $\hat w$ and $\hat v$ are real mixing matrices, subject
to the orthonormality relations

$${\sum_{\sss k=1}^{\sss n+m}}{u_{\sss{ik}}}{u_{\sss{jk}}}=
{\sum_{\sss k=1}^{\sss m}}{w_{\sss{ik}}}{w_{\sss{jk}}}=
{\sum_{\sss k=m+1}^{\sss m+n+o}}{v_{\sss{ik}}}{v_{\sss{jk}}}={\delta_{ij}}.
\EQN orthonorm$$ 
The dimension of the most general representation is ${\it
Dim}=2(2n+3m)+o$ where $n$ is the number of ${\bf(2,2)}$
representations, $m$ is the number of ${\bf(1,3)}\oplus {\bf(3,1)}$
representations, and $o$ is the number of ${\bf(1,1)}$
representations.  We work with $Dim=finite$ but it should be
understood that strictly speaking $Dim=\infty$ in large-$\nc$ QCD.

It is convenient to group the pion, the axialvector states and the
pseudoscalar states into the $n+m$-component vector $\Pi_i =(\pi,
A_{\sss 2}...A_{\sss l}, P_{\sss l+1}...P_{\sss n+m})$.  We can
then express the pion multiplet in the compact form:

\offparens 
$$\EQNalign{ 
&\ket{\Pi_a}_i=
{\sum_{\sss j=1}^{\sss m}}{u_{\sss{ij}}}{\ket{{\bf{A}}_a}_j}+
{\sum_{\sss j=m+1}^{\sss n+m}}{u_{\sss{ij}}}{\ket{{\bf{(2,2)}}_a}_j}
\qquad\qquad i=1\ldots (m+n)
\EQN genrep;a\cr
&\ket{{V}_a}_{i}=
{\sum_{\sss j=1}^{\sss m}}{w_{\sss{ij}}}{\ket{{\bf{V}}_a}_j}
\qquad\qquad\qquad\qquad\qquad\;\;
\qquad i=1\ldots m
\EQN genrep;b\cr
&\ket{S}_i=
{\sum_{\sss j=m+1}^{\sss n+m}}{v_{\sss{ij}}}{\ket{{\bf{(2,2)}}_{\sss 0}}_j}+
{\sum_{\sss j=n+m+1}^{\sss m+n+o}}{v_{\sss{ij}}}{\ket{{\bf{(1,1)}}}_j}
\qquad i=1\ldots (n+o).
\EQN genrep;c\cr}
$$ 
\autoparens
The helicity label has been supressed for ease of notation.  It will
be reintroduced below.  The pion and vector meson decay constants 
generalized from \Eq{infdecayone} are given by

$$\EQNalign{ 
&\bra{0}{{\cal A}_a}\ket{{\Pi}_b}_{i}={\delta_{ab}}{F_{\sss \Pi_i}}
\EQN decgenrel;a\cr
&\bra{0}{{\cal V}_a}\ket{{V}_b}_{i}={\delta_{ab}}{F_{\sss V_i}}
\EQN decgenrel;b\cr}
$$ 
and we now define the matrix elements of definite chirality:

\offparens
$$\EQNalign{ 
&\bra{0}{{\cal V}_a}{\ket{{{\bf V}_b}}_j}=
\bra{0}{{\cal A}_a}{\ket{{{\bf A}_b}}_j}
\equiv {\delta_{ab}}{F_j} \EQN nredecaydef;a\cr
&\bra{0}{{\cal V}_a}{\ket{{{\bf A}_b}}_j}=
\bra{0}{{\cal A}_a}{\ket{{{\bf V}_b}}_j}=0.
\EQN nredecaydef;b\cr}
$$ 
Now using \Eq{genrep}, \Eq{decgenrel} and \Eq{nredecaydef} gives

$$\EQNalign{ 
&{F_{\sss \Pi_i}}={\sum_{\sss j=1}^{\sss m}}{u_{\sss ij}}{F_j}
\EQN sf1decay;a\cr
&{F_{\sss V_i}}={\sum_{\sss j=1}^{\sss m}}{w_{\sss ij}}{F_j}.
\EQN sf1decay;b\cr}
$$
Of course these decay constants are nonvanishing only for pion and
vector meson states while the sums over states span all spins
according to \Tbl{gl}. The axialvector couplings are defined by

$$\EQNalign{ 
&{_i}\bra{\Pi_b}{Q^{\sss 5}_{a}}{\ket{S}_j}=
-i{\delta_{ab}}{G_{\sss {S_j}{\Pi_i}}}/{F_\pi}
\EQN coupgenrel;a\cr
&{_i}\bra{\Pi_b}{Q^{\sss 5}_{a}}{\ket{V_c}_j}=
-i{\epsilon_{abc}}{G_{\sss {V_j}{\Pi_i}}}/{F_\pi}.
\EQN coupgenrel;b\cr}
$$
From \Eq{gentensor}, \Eq{genrep} and \Eq{coupgenrel} it follows that

$$\EQNalign{
&{G_{\sss {S_j}{\Pi_i}}}/{F_\pi}=
-{\sum_{\sss k=m+1}^{\sss m+n}}{u_{\sss ik}}{v_{\sss jk}} \EQN picouple;a\cr
&{G_{\sss {V_j}{\Pi_i}}}/{F_\pi}=
{\sum_{\sss k=1}^{\sss m}}{u_{\sss ik}}{w_{\sss jk}}. \EQN picouple;b\cr}
$$
Note that the axial couplings are completely determined by the mixing
matrices. It is now easy to find relations that are independent of
mixing matrices by contracting \Eq{sf1decay} and \Eq{picouple}.

\vskip0.1in
\noindent {\twelvepoint{\it 3.4\quad Chiral Symmetry Constraints}}

There is one relation involving decay constants only:

$$
{\sum_{\sss i=1}^{\sss m+n}}{F_{\sss {\Pi_i}}}{F_{\sss {\Pi_i}}}=
{\sum_{\sss i=1}^{\sss m}}{F_{\sss {V_i}}}{F_{\sss {V_i}}}.
\EQN newmainaw1$$
There are three relations which contract one decay constant with
one axial coupling:

$$\EQNalign{ 
&{\sum_{\sss i=1}^{\sss m}}{F_{\sss {V_i}}}{G_{\sss {V_i}{\Pi_k}}}=
{F_\pi}{F_{\sss \Pi_k}} \EQN mainaw1;a\cr
&{\sum_{\sss j=1}^{\sss m+n}}{F_{\sss {\Pi_j}}}{G_{\sss {V_i}{\Pi_j}}}=
{F_\pi}{F_{\sss {V_i}}} \EQN mainaw1;b\cr
&{\sum_{\sss j=1}^{\sss m+n}}{F_{\sss {\Pi_j}}}{G_{\sss {S_i}{\Pi_j}}}=0.
\EQN mainaw1;c\cr}
$$
There are three relations which contract two axial couplings:

$$\EQNalign{ 
&{\sum_{\sss j=m+1}^{\sss m+n+o}}{G_{\sss {S_j}{\Pi_i}}}
{G_{\sss {S_j}{\Pi_k}}}+
{\sum_{\sss j=1}^{\sss m}}{G_{\sss {V_j}{\Pi_i}}}{G_{\sss {V_j}{\Pi_k}}}=
\yo1{\delta_{ik}} \EQN mainaw1as;a\cr
&{\sum_{\sss i=1}^{\sss m+n}}{G_{\sss {V_j}{\Pi_i}}}{G_{\sss {V_k}{\Pi_i}}}=
\yo1{\delta_{jk}} \EQN mainaw1as;b\cr
&{\sum_{\sss i=1}^{\sss m+n}}{G_{\sss {S_j}{\Pi_i}}}{G_{\sss {V_k}{\Pi_i}}}=0.
\EQN mainaw1as;c\cr}
$$
These are the basic sum rules. (In an appendix these sum rules are
derived directly from the chiral algebra.) They can be further
contracted with axial couplings and decay constants to give other sum
rules. In compact notation the complete set of sum rules which follow
from
\Eq{newmainaw1}-\Eq{mainaw1as} is:

\vskip0.2in

\noindent{\underbar {\it Two-point functions (SFSR1):}}

$$
{\sum_{\sss V}}{F_{\sss V}^2}-{\sum_{\sss
A}}{F_{\sss A}^2}= {F_\pi^2} 
\EQN sfsroneas$$

\vskip0.2in

\noindent{\underbar {\it Three-point functions (one external current):}}

$$\EQNalign{ 
&{\sum_{\sss V}}{F_{\sss V}}{G_{{\sss V}\pi}}=\yo1
\EQN eckerszs;a\cr
&{\sum_{\sss V}}{F_{\sss V}}{G^{\sss ({0})}_{\sss V{A_i}}}=
{F_\pi}{F_{\sss {A_i}}}
\EQN eckerszs;b\cr
&{\sum_{\sss V}}{F_{\sss V}}{G^{\sss ({0})}_{\sss V{P_i}}}=0
\EQN eckerszs;c\cr
&{\sum_{\sss A}}{F_{\sss A}}{G^{\sss ({0})}_{\sss {V_i}{A}}}+
{F_{\pi}}{G_{{\sss {V_i}}\pi}}={F_\pi}{F_{\sss {V_i}}}
\EQN eckerszs;d\cr
&{\sum_{\sss A}}{F_{\sss A}}{G^{\sss ({0})}_{\sss {S_i}{A}}}+
{F_{\pi}}{G_{{\sss {S_i}}\pi}}=0
\EQN eckerszs;e\cr}
$$

\figure{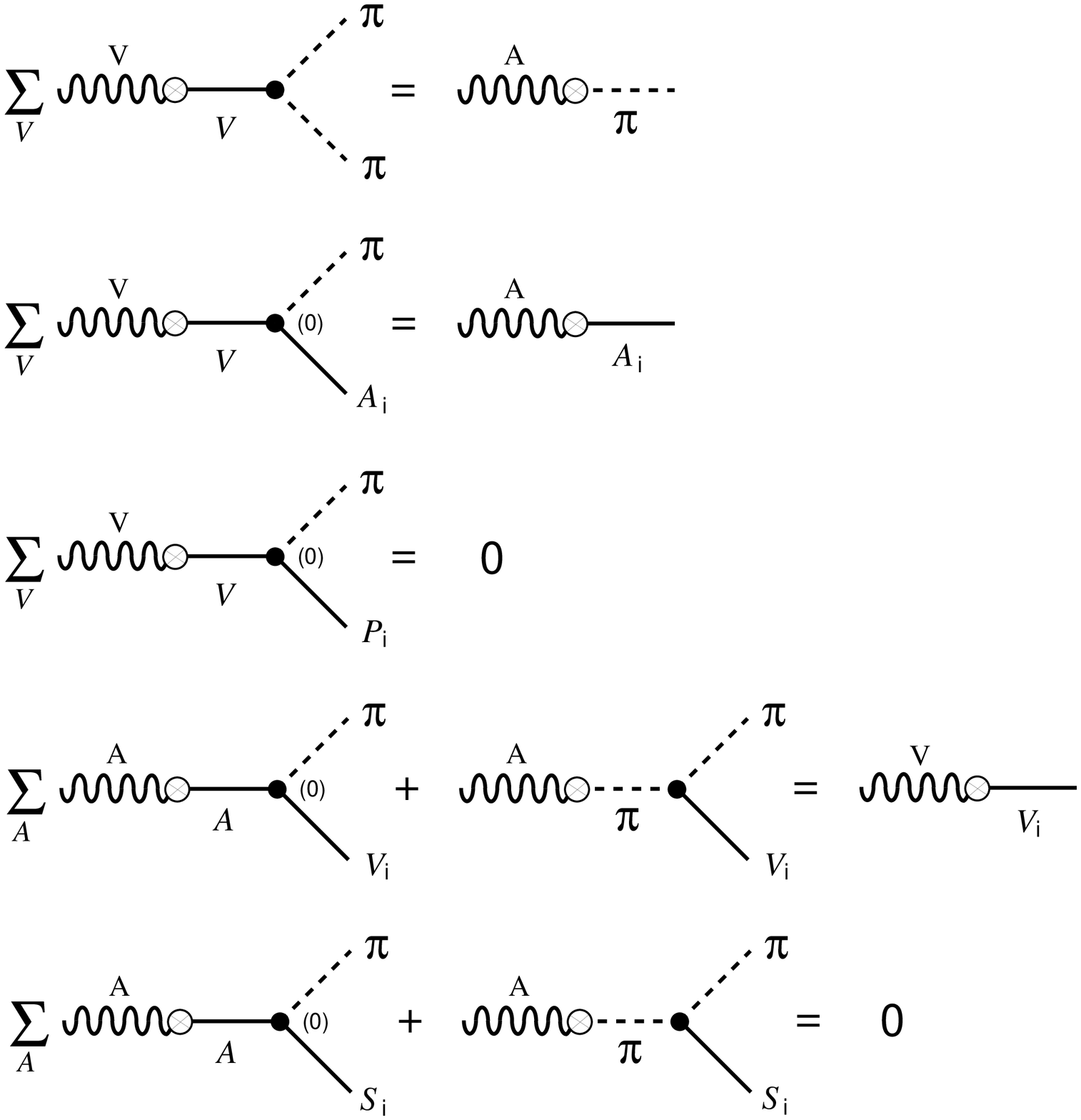}
\epsfxsize 3.6in 
\centerline{\epsfbox{threecurrents.eps}}
\caption{Sum rules for three-point functions with
one external current (\Eq{eckerszs}).  The uppermost sum rule
constrains the pion vector form factor.}
\endcaption
\endfigure

\vskip0.2in

\noindent{\underbar {\it Three-point functions (two external currents):}}

$$
{\sum_{\sss V,A}}{F_{\sss V}}{G^{\sss ({0})}_{\sss {V}{A}}}{F_{\sss A}}=
{F_\pi}{\sum_{\sss A}}{F_{\sss A}^2}
\EQN eckerszs2$$

\vskip0.2in

\noindent{\underbar {\it Four-point functions 
(no external currents):}}

\offparens
$$\EQNalign{ 
&{\sum_{\sss S}}{G^2_{{\sss S}\pi}}+{\sum_{\sss V}}{G^2_{{\sss V}\pi}}=\yo1
\EQN aw1comp2;a\cr
&{\sum_{\sss S}}{G_{{\sss S}\pi}}{G^{\sss ({0})}_{\sss S{A_i}}}
+{\sum_{\sss V}}{G_{{\sss V}\pi}}{G^{\sss ({0})}_{\sss V{A_i}}}=0
\EQN aw1comp2;b\cr
&{\sum_{\sss S}}{G_{{\sss S}\pi}}{G^{\sss ({0})}_{\sss S{P_i}}}
+{\sum_{\sss V}}{G_{{\sss V}\pi}}{G^{\sss ({0})}_{\sss V{P_i}}}=0
\EQN aw1comp2;c\cr
&{\sum_{\sss S}}{G^{\sss ({0})}_{\sss S{P_i}}}{G^{\sss ({0})}_{\sss S{A_k}}}+
{\sum_{\sss V}}{G^{\sss ({0})}_{\sss V{P_i}}}{G^{\sss ({0})}_{\sss V{A_k}}}=0
\EQN aw1comp2;d\cr
&{\sum_{\sss S}}{G^{\sss ({0})}_{\sss S{A_i}}}{G^{\sss ({0})}_{\sss S{A_k}}}+
{\sum_{\sss V}}{G^{\sss ({0})}_{\sss V{A_i}}}{G^{\sss ({0})}_{\sss V{A_k}}}=
\yo1{\delta_{ik}}
\EQN aw1comp2;e\cr
&{\sum_{\sss S}}{G^{\sss ({0})}_{\sss S{P_i}}}{G^{\sss ({0})}_{\sss S{P_k}}}+
{\sum_{\sss V}}{G^{\sss ({0})}_{\sss V{P_i}}}{G^{\sss ({0})}_{\sss V{P_k}}}=
\yo1{\delta_{ik}}
\EQN aw1comp2;f\cr
&{G_{{\sss {V_i}}\pi}}{G_{{\sss {V_k}}\pi}}
+{\sum_{\sss A}}{G^{\sss ({0})}_{\sss {V_i}A}}{G^{\sss ({0})}_{\sss {V_k}A}}
+{\sum_{\sss P}}{G^{\sss ({0})}_{\sss {V_i}P}}{G^{\sss ({0})}_{\sss {V_k}P}}
=\yo1{\delta_{ik}}
\EQN aw1comp2;g\cr
&{G_{{\sss {S_i}}\pi}}{G_{{\sss {V_k}}\pi}}
+{\sum_{\sss A}}{G^{\sss ({0})}_{\sss {S_i}A}}{G^{\sss ({0})}_{\sss {V_k}A}}
+{\sum_{\sss P}}{G^{\sss ({0})}_{\sss {S_i}P}}{G^{\sss ({0})}_{\sss {V_k}P}}
=0\EQN aw1comp2;h\cr}
$$

\vskip0.2in

\noindent{\underbar {\it Four-point functions (one external current):}}

\offparens
$$\EQNalign{ 
&{\sum_{\sss A,S}}{F_{\sss A}}{G^{\sss ({0})}_{\sss {S}A}}{G_{{\sss S}\pi}}+
{\sum_{\sss A,V}}{F_{\sss A}}{G^{\sss ({0})}_{\sss {V}A}}{G_{{\sss V}\pi}}
=0 \EQN aw1comp2contr1;a\cr
&{\sum_{\sss A,S}}{F_{\sss A}}{G^{\sss ({0})}_{\sss {S}A}}
{G^{\sss ({0})}_{\sss S{A_i}}}+
{\sum_{\sss A,V}}{F_{\sss A}}{G^{\sss ({0})}_{\sss {V}A}}
{G^{\sss ({0})}_{\sss V{A_i}}}
=\yo1 {F_{\sss A_i}}
\EQN aw1comp2contr1;b\cr
&{\sum_{\sss A,S}}{F_{\sss A}}{G^{\sss ({0})}_{\sss {S}A}}
{G^{\sss ({0})}_{\sss S{P_i}}}+
{\sum_{\sss A,V}}{F_{\sss A}}{G^{\sss ({0})}_{\sss {V}A}}
{G^{\sss ({0})}_{\sss V{P_i}}}=0
\EQN aw1comp2contr1;c\cr
&{\sum_{\sss V}}{F_{\sss V}}{G_{{\sss V}\pi}}{G_{{\sss {V_i}}\pi}}+
{\sum_{\sss V,A}}{F_{\sss V}}
{G^{\sss ({0})}_{\sss {V}A}}{G^{\sss ({0})}_{\sss {V_i}A}}+
{\sum_{\sss V,P}}{F_{\sss V}}
{G^{\sss ({0})}_{\sss {V}P}}{G^{\sss ({0})}_{\sss {V_i}P}}
=\yo1 {F_{\sss V_i}}
\EQN aw1comp2contr1;d\cr
&{\sum_{\sss V}}{F_{\sss V}}{G_{{\sss {V}}\pi}}{G_{{\sss {S_k}}\pi}}
+{\sum_{\sss V,A}}{F_{\sss V}}{G^{\sss ({0})}_{\sss {V}A}}
{G^{\sss ({0})}_{\sss {S_k}A}}
+{\sum_{\sss V,P}}{F_{\sss V}}{G^{\sss ({0})}_{\sss {V}P}}
{G^{\sss ({0})}_{\sss {S_k}P}}
=0
\EQN aw1comp2contr1;e\cr}
$$

\vskip0.2in

\figure{threecurrentsb.eps}
\epsfxsize 3.7in 
\centerline{\epsfbox{threecurrentsb.eps}}
\caption{Sum rule for three-point functions with two external currents
(\Eq{eckerszs2}).} 
\endcaption
\endfigure

\noindent{\underbar {\it Four-point functions (two external currents):}}

\offparens 
$$\EQNalign{ 
&{\sum_{\sss A,S,A'}} {F_{\sss A}}{G^{\sss
({0})}_{\sss {S}A}} {G^{\sss ({0})}_{\sss {S}A'}}{F_{\sss A'}}+
{\sum_{\sss A,V,A'}} {F_{\sss A}}{G^{\sss ({0})}_{\sss {V}A}} {G^{\sss
({0})}_{\sss {V}A'}}{F_{\sss A'}}= \yo1{\sum_{\sss A}}{F_{\sss A}^2}
\EQN aw1comp2contr2;a\cr 
&{\sum_{\sss V,V'}}{F_{\sss V}}{G_{{\sss
V}\pi}}{G_{{\sss V'}\pi}}{F_{\sss V'}} +{\sum_{\sss V,A,V'}}{F_{\sss
V}} {G^{\sss ({0})}_{\sss {V}A}} {G^{\sss ({0})}_{\sss {V'}A}}{F_{\sss
V'}} +{\sum_{\sss V,P,V'}}{F_{\sss V}} {G^{\sss ({0})}_{\sss {V}P}}
{G^{\sss ({0})}_{\sss {V'}P}}{F_{\sss V'}}= \yo1{\sum_{\sss
V}}{F_{\sss V}^2}. \EQN aw1comp2contr2;b\cr} $$ 
We have reintroduced the helicity label to denote those couplings
which generally have nonzero helicity components which are not
constrained by the pion representation. The first sum rule is the
first spectral function sum rule, {\it SFSR1}. Here we see that there
are an infinite number of additional sum rules which are at precisely
the same level of theoretical rigor. The new sum rules are illustrated
diagramatically in \Fig{threecurrents.eps} (\Eq{eckerszs}),
\Fig{threecurrentsb.eps} (\Eq{eckerszs2}), \Fig{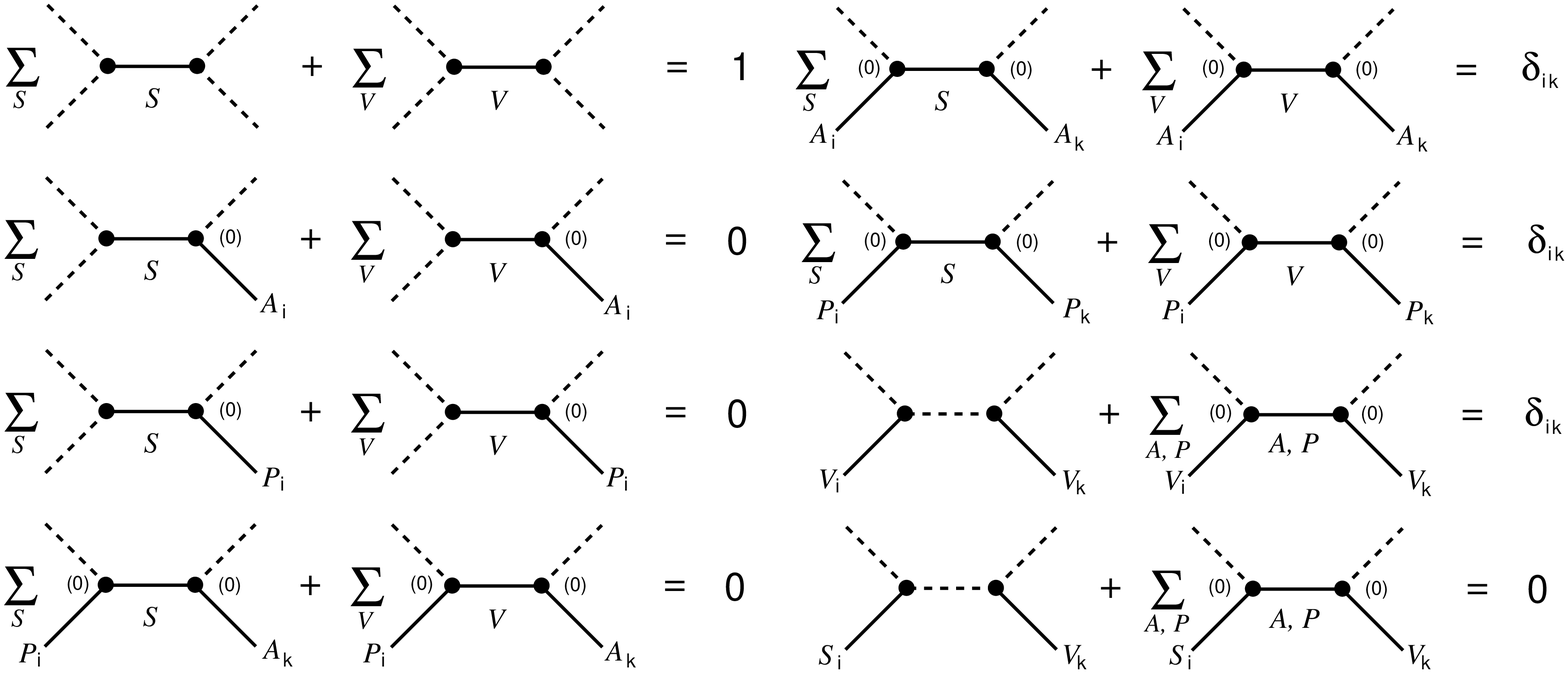}
(\Eq{aw1comp2}), \Fig{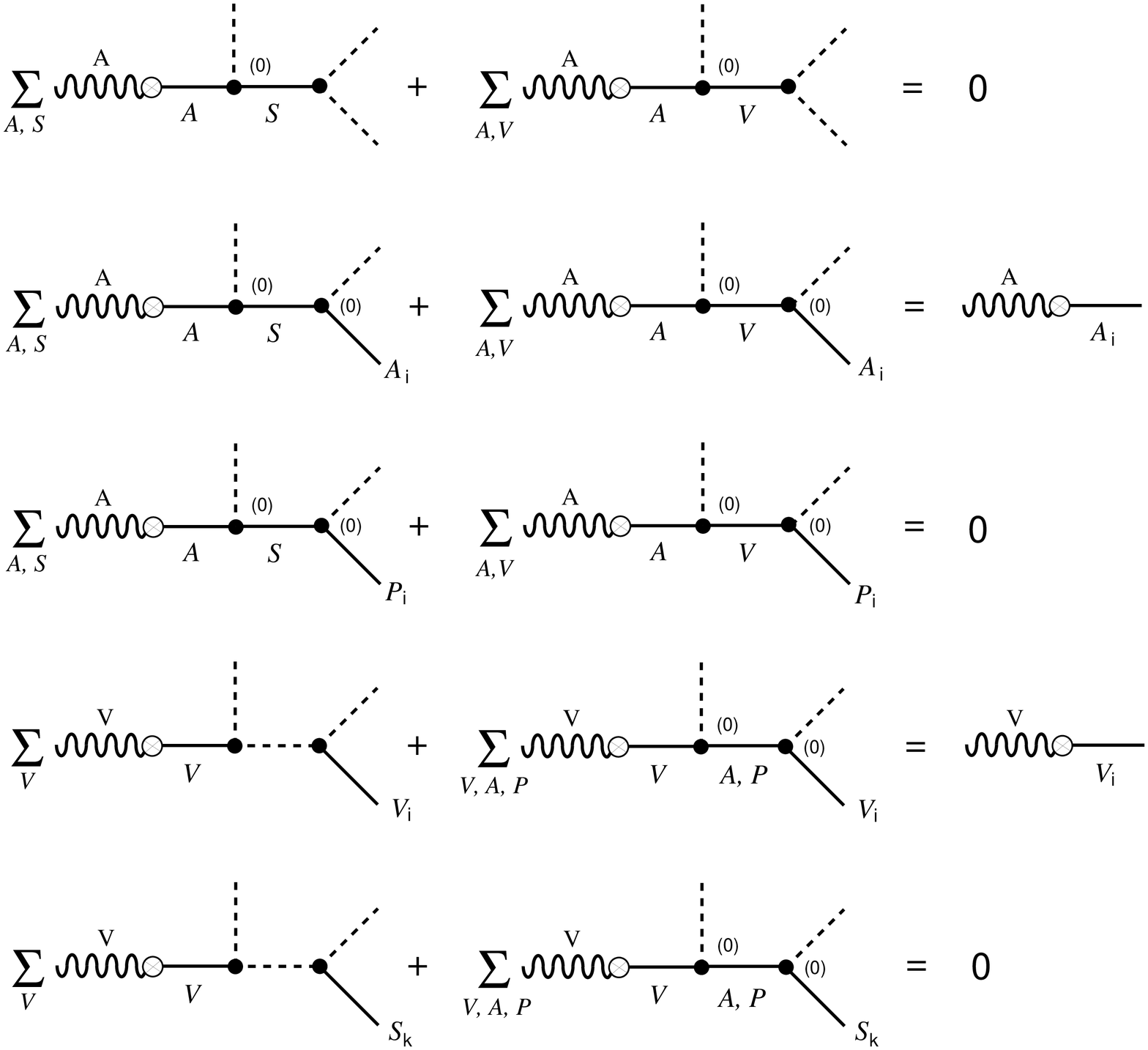} (\Eq{aw1comp2contr1}) and
\Fig{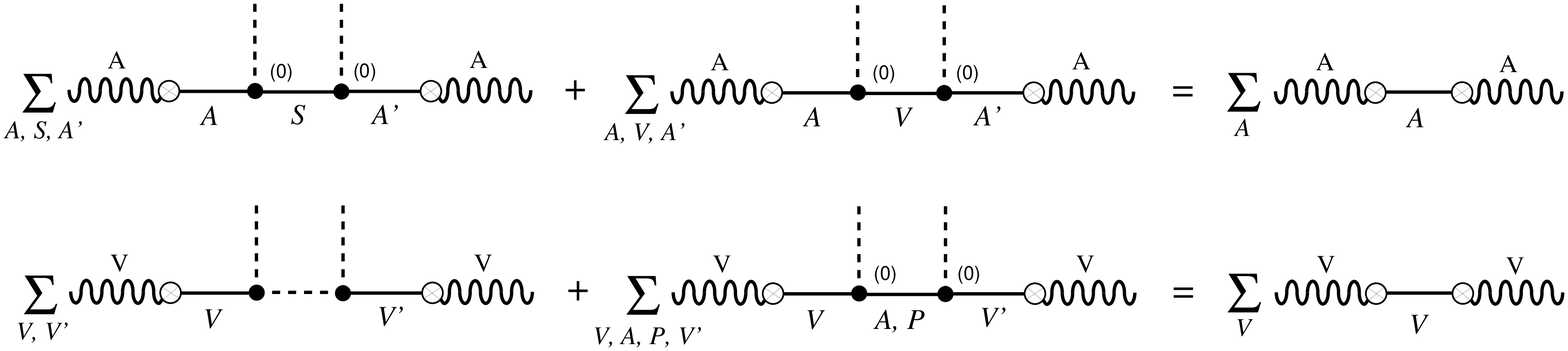} (\Eq{aw1comp2contr2}).

\figure{modified2.eps}
\epsfxsize 6in
\centerline{\epsfbox{modified2.eps}}
\caption{Sum rules for four-point functions with no external currents
(\Eq{aw1comp2}). Dotted lines are pions.}
\endcaption
\endfigure

\vskip0.1in
\noindent {\twelvepoint{\it 3.5\quad Symmetry Breaking Constraints}}

Given that the chiral charges annihilate the vacuum at infinite
momentum, one may wonder how symmetry breaking manifests itself. Of
course it must be present in order to split states within chiral
multiplets.  At infinite momentum spontaneous symmetry breaking
implies

\offparens 
$$ [{Q^{5}_{a}},{{\hat {\cal H}}_{\sss QCD}^\infty}] \neq 0
\EQN hamil $$ \autoparens 
where ${{\hat {\cal H}}_{\sss QCD}^\infty}$ is the QCD Hamiltonian at
infinite momentum.  This is simply the statement ---familiar to
practitioners of light-front field theory--- that at infinite
momentum, the effects of spontaneous symmetry breaking appear as {\it
explicit} breaking terms in the Hamiltonian. At infinite momentum the
Hamiltonian can be formally expanded as:

\figure{newfourcurrentsb.eps}
\epsfxsize 4.5in
\centerline{\epsfbox{newfourcurrentsb.eps}}
\caption{Sum rules for four-point functions with one external current
(\Eq{aw1comp2contr1}). Dotted lines are pions.}
\endcaption
\endfigure

\offparens
$$
{{\hat {\cal H}}_{\sss QCD}^\infty}=
\sqrt{{{\hat P}^2} + {{\hat M}^2}}=
|{{\hat P}}| + {{{\hat M}^2}\over{2|{{\hat P}}|}}
+O(|{{\hat P}}|^{-3}).
\EQN hamilexpan
$$
\autoparens 
Therefore, spontaneous chiral symmetry breaking implies

\offparens
$$
[{Q^{5}_{a}},{{\hat {M}}^2}]\neq 0
\EQN hamilmass
$$
where ${{\hat {M}}^2}$ is the hadronic mass-squared matrix.  The
general solution to \Eq{hamilmass} is

$$
{{\hat M}^2}={{\hat M}^2_{\sss {\bf{(1,1)}}}}+
{\sum_{\sss {\cal R}}}{{\hat M}^2_{\sss {\cal R}}}
\EQN genmasssqmat$$
where $\bf{(1,1)}$ is of course the singlet representation and ${\cal
R}$ is a nontrivial $SU({2})\times SU({2})$ representation. There is
no sense in which ${{\hat M}^2_{\sss {\cal R}}}$ is small.  From the
allowed representations for the states it is straightforward to show
that ---without loss of generality--- the most general symmetry
breaking mass-squared matrix can be written as

$$
{\sum_{\sss {\cal R}}}{{\hat M}^2_{\sss {\cal R}}}=
{{\hat M}^2_{\sss {\bf{(2,2)}}}}+
{{\hat M}^2_{\sss {\bf{(3,3)}}}}.
\EQN twostsol$$
The matrix elements of ${{\hat M}^2_{\sss \bf{(1,1)}}}$ between the
states of definite chirality are defined as

\offparens
$$\EQNalign{ 
&{_i}\bra{{\bf{V}}_a}
{{\hat M}^2_{\sss \bf{(1,1)}}}
\ket{{\bf{V}}_b}{_j}=
{_i}\bra{{\bf{A}}_a}
{{\hat M}^2_{\sss \bf{(1,1)}}}
\ket{{\bf{A}}_b}{_j}\equiv
{\delta_{ab}}{m^2_{ij}}
\EQN massesoneone;a\cr
&{_i}\bra{{\bf{(2,2)}}_\alpha}
{{\hat M}^2_{\sss \bf{(1,1)}}}
\ket{{\bf{(2,2)}}_\beta}{_j}\equiv
{\delta_{\alpha\beta}}{n^2_{ij}}
\EQN massesoneone;b\cr
&{_i}\bra{{\bf{(1,1)}}}
{{\hat M}^2_{\sss \bf{(1,1)}}}
\ket{{\bf{(1,1)}}}{_j}\equiv
{o^2_{ij}}.
\EQN massesoneone;c\cr}
$$ 
The matrix elements of ${{\hat M}^2_{\sss \bf{(2,2)}}}$ 
are defined as

\offparens
$$\EQNalign{ 
&{_i}\bra{{\bf{A}}_a}
{{\hat M}^2_{\sss \bf{(2,2)}}}
\ket{{\bf{(2,2)}}_b}{_j}=
{_i}\bra{{\bf{(2,2)}}_a}
{{\hat M}^2_{\sss \bf{(2,2)}}}
\ket{{\bf{A}}_b}{_j}\equiv
{\delta_{ab}}{{\bar n}^2_{ij}}
\EQN massestwotwo;a\cr
&{_i}\bra{{\bf{(1,1)}}}
{{\hat M}^2_{\sss \bf{(2,2)}}}
\ket{{\bf{(2,2)}}_0}{_j}=
{_i}\bra{{\bf{(2,2)}}_0}
{{\hat M}^2_{\sss \bf{(2,2)}}}
\ket{{\bf{(1,1)}}}{_j}\equiv
{{\bar o}^2_{ij}}.
\EQN massestwotwo;b}
$$
The matrix elements of ${{\hat M}^2_{\sss (3,3)}}$ 
are defined as

\offparens
$$
-{_i}\bra{{\bf{V}}_a}
{{\hat M}^2_{\sss \bf{(3,3)}}}
\ket{{\bf{V}}_b}{_j}=
{_i}\bra{{\bf{A}}_a}
{{\hat M}^2_{\sss \bf{(3,3)}}}
\ket{{\bf{A}}_b}{_j}\equiv
{\delta_{ab}}{{\bar m}^2_{ij}}.
\EQN massesthreethree$$

The physical masses of the pion multiplet states are

\offparens
$$\EQNalign{ 
&{_i}\bra{{\Pi}_a}{{\hat M}^2}\ket{{\Pi}_b}{_j}=
{\delta_{ij}}{\delta_{ab}}{M_{\sss {\Pi_i}}^2}
\EQN massesdef;a\cr
&{_i}\bra{V_a}{{\hat M}^2}\ket{V_b}{_j}=
{\delta_{ij}}{\delta_{ab}}{M_{\sss {V_i}}^2}
\EQN massesdef;b\cr
&{_i}\bra{S}{{\hat M}^2}\ket{S}{_j}=
{\delta_{ij}}{M_{\sss {S_i}}^2},
\EQN massesdef;c\cr}
$$ 
from which we obtain using \Eq{genrep}, \Eq{massesoneone}, \Eq{massestwotwo},
and \Eq{massesthreethree}:

\offparens
$$\EQNalign{ 
&{\delta_{ij}}{M_{\sss {\Pi_i}}^2}=
{\sum_{\sss k=1}^{\sss m}}{\sum_{\sss l=1}^{\sss m}}
{u_{\sss ik}}{u_{\sss jl}}\lbrace
{{m}^2_{kl}}+{{\bar m}^2_{kl}}\rbrace +
{\sum_{\sss k=m+1}^{\sss m+n}}{\sum_{\sss l=m+1}^{\sss m+n}}
{u_{\sss ik}}{u_{\sss jl}}{{n}^2_{kl}}
+2{\sum_{\sss k=1}^{\sss m}}{\sum_{\sss l=m+1}^{\sss m+n}}
{u_{\sss ik}}{u_{\sss jl}}{{\bar n}^2_{kl}}\qquad
\EQN physmasses;a\cr
&{\delta_{ij}}{M_{\sss {V_i}}^2}=
{\sum_{\sss k=1}^{\sss m}}{\sum_{\sss l=1}^{\sss m}}
{w_{\sss ik}}{w_{\sss jl}}\lbrace
{{m}^2_{kl}}-{{\bar m}^2_{kl}}\rbrace
\EQN physmasses;b\cr
&{\delta_{ij}}{M_{\sss {S_i}}^2}=
{\sum_{\sss k=m+1}^{\sss m+n}}{\sum_{\sss l=m+1}^{\sss m+n}}
{v_{\sss ik}}{v_{\sss jl}}{{n}^2_{kl}}+
{\sum_{\sss k=n+m+1}^{\sss m+n+o}}{\sum_{\sss l=n+m+1}^{\sss m+n+o}}
{v_{\sss ik}}{v_{\sss jl}}{{o}^2_{kl}}+
2{\sum_{\sss k=m+1}^{\sss m+n}}{\sum_{\sss l=n+m+1}^{\sss m+n+o}}
{v_{\sss ik}}{v_{\sss jl}}{{\bar o}^2_{kl}}.
\EQN physmasses;c\cr}
$$

\figure{newfourcurrentsc.eps}
\epsfxsize 5.5in
\centerline{\epsfbox{newfourcurrentsc.eps}}
\caption{Sum rules for four-point functions with two external currents
(\Eq{aw1comp2contr2}). Dotted lines are pions.}
\endcaption
\endfigure

Contracting with the decay constants of \Eq{sf1decay} it is then easy to obtain
the sum rule:

$$
{\sum_{\sss i=1}^{\sss m}}{\sum_{\sss j=1}^{\sss m}}
{F_{\sss {V_i}}}{F_{\sss {V_j}}}
{\delta_{ij}}{M_{\sss {V_i}}^2}-
{\sum_{\sss i=1}^{\sss m+n}}{\sum_{\sss j=1}^{\sss m+n}}
{F_{\sss {\Pi_i}}}{F_{\sss {\Pi_j}}}
{\delta_{ij}}{M_{\sss {\Pi_i}}^2}=
-2{\sum_{\sss k=1}^{\sss m}}{\sum_{\sss l=1}^{\sss m}}
{F_k}{F_l}{{\bar m}^2_{kl}}.
\EQN sfsr1lzero$$
We also have the relation

\offparens
$$
{\sum_{\sss i=1}^{\sss n+m}}{\sum_{\sss j=1}^{\sss n+m}}
{F_{\sss {\Pi_i}}}{F_{\sss {\Pi_j}}}
{_i}\bra{{\Pi}_a}{{\hat M}^2_{\sss \bf{(3,3)}}}\ket{{\Pi}_b}{_j}=
-{\sum_{\sss i=1}^{\sss m}}{\sum_{\sss j=1}^{\sss m}}
{F_{\sss {V_i}}}{F_{\sss {V_j}}}
{_i}\bra{{V}_a}{{\hat M}^2_{\sss \bf{(3,3)}}}\ket{{V}_b}{_j}=
{\delta_{ab}}{\sum_{\sss k=1}^{\sss m}}{\sum_{\sss l=1}^{\sss m}}
{F_k}{F_l}{{\bar m}^2_{kl}}.
\EQN msqrelation$$\autoparens
In compact notation, combining \Eq{sfsr1lzero} and \Eq{msqrelation},
we have

$$
{\sum_{\sss V}}{F_{\sss V}^2}{M_{\sss V}^{2}}-
{\sum_{\sss A}}{F_{\sss A}^2}{M_{\sss A}^{2}}=
-2{\sum_{\sss \Pi}}{\sum_{\sss \Pi'}}
{F_{\sss {\Pi}}}{F_{\sss {\Pi'}}}
\bra{\Pi}{{\hat M}^2_{\sss \bf{(3,3)}}}\ket{\Pi'}.
\EQN sfsr1lzerocompact$$
Comparing with the OPE in \Eq{presfsrex;b} gives

$$
{\vev{{\cal O}}^{\sss d=4}_{\sss \bf{(3,3)}}}=
2{\sum_{\sss \Pi}}{\sum_{\sss \Pi'}}
{F_{\sss {\Pi}}}{F_{\sss {\Pi'}}}
\bra{\Pi}{{\hat M}^2_{\sss \bf{(3,3)}}}\ket{\Pi'}.
\EQN opetomsq$$
Of course this OPE coefficient vanishes in 
QCD. Therefore ${{\hat M}^2_{\sss \bf{(3,3)}}}=0$ and

$$
{{\hat M}^2}={{\hat M}^2_{\sss {\bf{(1,1)}}}}+{{\hat M}^2_{\sss {\bf{(2,2)}}}}
\EQN genmasssqmatqcd$$
in large-$\nc$ QCD.  This chiral decomposition of the mass-squared
matrix therefore implies {\it SFSR2},

$$
{\sum_{\sss V}}{F_{\sss V}^2}{M_{\sss V}^{2}}-
{\sum_{\sss A}}{F_{\sss A}^2}{M_{\sss A}^{2}}=0.
\EQN sfsr2yetagain$$
It is interesting that one can use the OPE to constrain the form of
the meson mass-squared matrix.  This result justifies the assumption
of Coleman and Witten that the order parameter of chiral symmetry
breaking transforms purely as $\bf{(2,2)}$ in the large-$\nc$
limit\ref{cole}.

It is now straightforward to find new relations that are independent of
mixing matrices by contracting \Eq{picouple} with
\Eq{physmasses}:

\offparens
$$\EQNalign{ 
&{\sum_{\sss j=m+1}^{\sss m+n+o}}{G_{\sss {S_j}{\Pi_i}}}
{G_{\sss {S_j}{\Pi_k}}}
({M_{\sss \Pi_i}^{2}}+{M_{\sss \Pi_k}^{2}}-2{M_{\sss S_j}^{2}})\cr
&-{\sum_{\sss j=1}^{\sss m}}{G_{\sss {V_j}{\Pi_i}}}
{G_{\sss {V_j}{\Pi_k}}}
({M_{\sss \Pi_i}^{2}}+{M_{\sss \Pi_k}^{2}}-2{M_{\sss V_j}^{2}})=0
\EQN genmasssumsol;a\cr
&{\sum_{\sss k=1}^{\sss m+n}}{G_{\sss {V_i}{\Pi_k}}}
{G_{\sss {V_j}{\Pi_k}}}
({M_{\sss V_i}^{2}}+{M_{\sss V_j}^{2}}-2{M_{\sss \Pi_k}^{2}})=0
\EQN genmasssumsol;b\cr
&{\sum_{\sss k=1}^{\sss m+n}}{G_{\sss {S_i}{\Pi_k}}}
{G_{\sss {V_j}{\Pi_k}}}
({M_{\sss V_j}^{2}}-{M_{\sss S_i}^{2}})=0.
\EQN genmasssumsol;c\cr}
$$ 
The last sum rule is trivially satisfied via \Eq{mainaw1as;c}.  (In an
appendix these sum rules are derived directly from the chiral
algebra.) It is also straightforward to relate the axial couplings
and masses to the symmetry breaking parameters. For instance, one finds

\offparens
$$
{_i}\bra{{\Pi}_a}
{{\hat M}^2_{\sss \bf{(2,2)}}}\ket{{\Pi}_b}{_j}\;\yo1
={\delta_{ab}}\lbrack{\sum_{\sss l=1}^{\sss m}}{G_{\sss {V_l}{\Pi_i}}}
{G_{\sss {V_l}{\Pi_j}}}
({M_{\sss \Pi_i}^{2}}+{M_{\sss \Pi_j}^{2}}-2{M_{\sss V_l}^{2}})\rbrack,
\EQN pomestuff
$$ 
a result which we will return to below.

In compact notation the complete set of mass-squared sum rules which
follow from \Eq{sfsr1lzerocompact} and \Eq{genmasssumsol} are:

\vskip0.2in

\noindent{\underbar {\it Two-point functions (SFSR2):}}

$$
{\sum_{\sss V}}{F_{\sss V}^2}{M_{\sss V}^{2}}-
{\sum_{\sss A}}{F_{\sss A}^2}{M_{\sss A}^{2}}=0
\EQN sfsr2compver$$

\figure{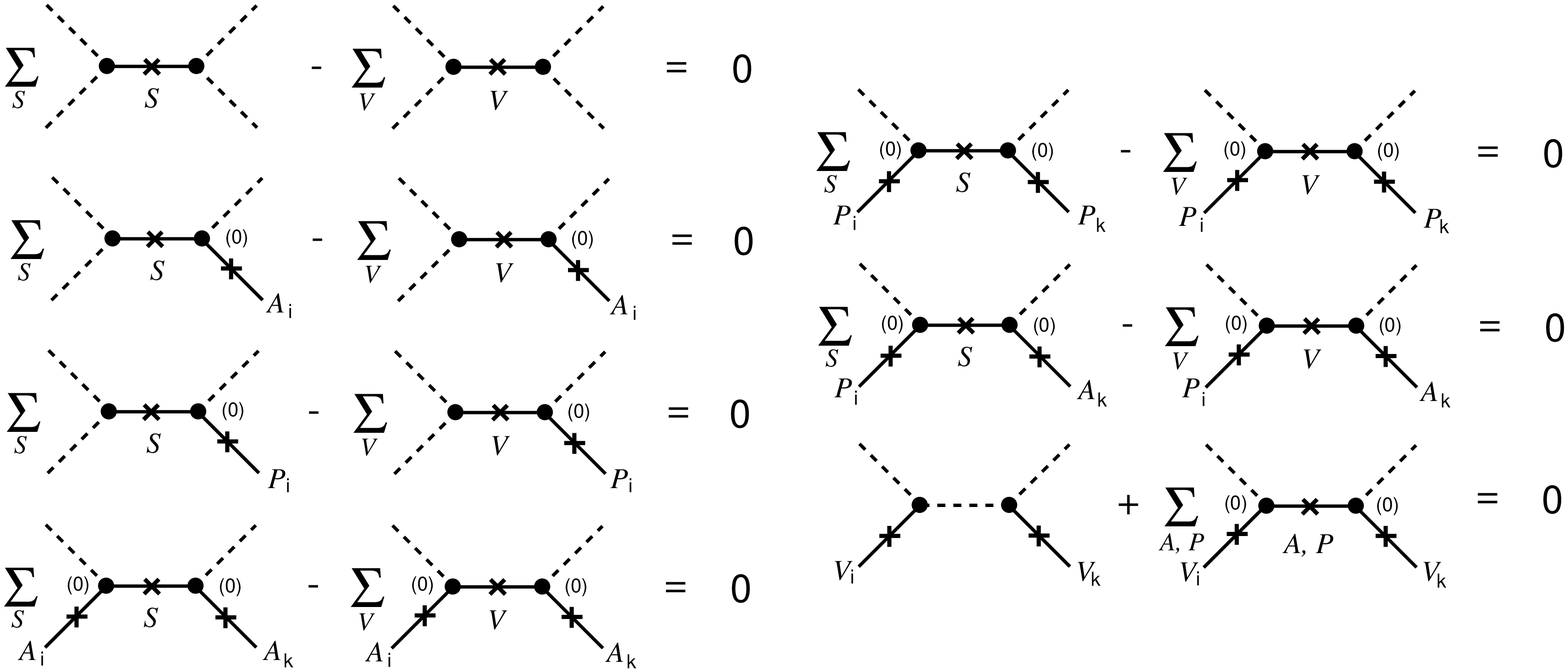}
\epsfxsize 6in
\centerline{\epsfbox{yetagain.eps}}
\caption{Sum rules for four-point functions with no external currents
(\Eq{sconcomp2}). Dotted lines are pions. Crosses represent
mass-squared insertions.}
\endcaption
\endfigure

\noindent{\underbar {\it Four-point functions (no external currents):}}

\offparens
$$\EQNalign{ 
&{\sum_{\sss S}}{G^2_{{\sss S}\pi}}{M_{\sss S}^{2}}-
{\sum_{\sss V}}{G^2_{{\sss V}\pi}}{M_{\sss V}^{2}}=0
\EQN sconcomp2;a\cr
&{\sum_{\sss S}}{G_{{\sss S}\pi}}{G^{\sss ({0})}_{\sss S{A_i}}}
({M_{\sss A_i}^{2}}-2{M_{\sss S}^{2}})
-{\sum_{\sss V}}{G_{{\sss V}\pi}}{G^{\sss ({0})}_{\sss V{A_i}}}
({M_{\sss A_i}^{2}}-2{M_{\sss V}^{2}})=0
\EQN sconcomp2;b\cr
&{\sum_{\sss S}}{G_{{\sss S}\pi}}{G^{\sss ({0})}_{\sss S{P_i}}}
({M_{\sss P_i}^{2}}-2{M_{\sss S}^{2}})
-{\sum_{\sss V}}{G_{{\sss V}\pi}}{G^{\sss ({0})}_{\sss V{P_i}}}
({M_{\sss P_i}^{2}}-2{M_{\sss V}^{2}})=0
\EQN sconcomp2;c\cr
&{\sum_{\sss S}}{G^{\sss ({0})}_{\sss S{A_i}}}{G^{\sss ({0})}_{\sss S{A_k}}}
({M_{\sss A_i}^{2}}+{M_{\sss A_k}^{2}}-2{M_{\sss S}^{2}})-
{\sum_{\sss V}}{G^{\sss ({0})}_{\sss V{A_i}}}{G^{\sss ({0})}_{\sss V{A_k}}}
({M_{\sss A_i}^{2}}+{M_{\sss A_k}^{2}}-2{M_{\sss V}^{2}})=0
\EQN sconcomp2;d\cr
&{\sum_{\sss S}}{G^{\sss ({0})}_{\sss S{P_i}}}{G^{\sss ({0})}_{\sss S{P_k}}}
({M_{\sss P_i}^{2}}+{M_{\sss P_k}^{2}}-2{M_{\sss S}^{2}})-
{\sum_{\sss V}}{G^{\sss ({0})}_{\sss V{P_i}}}{G^{\sss ({0})}_{\sss V{P_k}}}
({M_{\sss P_i}^{2}}+{M_{\sss P_k}^{2}}-2{M_{\sss V}^{2}})=0
\EQN sconcomp2;e\cr
&{\sum_{\sss S}}{G^{\sss ({0})}_{\sss S{P_i}}}{G^{\sss ({0})}_{\sss S{A_k}}}
({M_{\sss P_i}^{2}}+{M_{\sss A_k}^{2}}-2{M_{\sss S}^{2}})-
{\sum_{\sss V}}{G^{\sss ({0})}_{\sss V{P_i}}}{G^{\sss ({0})}_{\sss V{A_k}}}
({M_{\sss P_i}^{2}}+{M_{\sss A_k}^{2}}-2{M_{\sss V}^{2}})=0
\EQN sconcomp2;f\cr
&{G_{{\sss {V_i}}\pi}}{G_{{\sss {V_k}}\pi}}
({M_{\sss V_i}^{2}}+{M_{\sss V_k}^{2}})+
{\sum_{\sss A}}{G^{\sss ({0})}_{\sss {V_i}A}}{G^{\sss ({0})}_{\sss {V_k}A}}
({M_{\sss V_i}^{2}}+{M_{\sss V_k}^{2}}-2{M_{\sss A}^{2}})\cr
&+{\sum_{\sss P}}{G^{\sss ({0})}_{\sss {V_i}P}}{G^{\sss ({0})}_{\sss {V_k}P}}
({M_{\sss V_i}^{2}}+{M_{\sss V_k}^{2}}-2{M_{\sss P}^{2}})=0
\EQN sconcomp2;g\cr}
$$

\figure{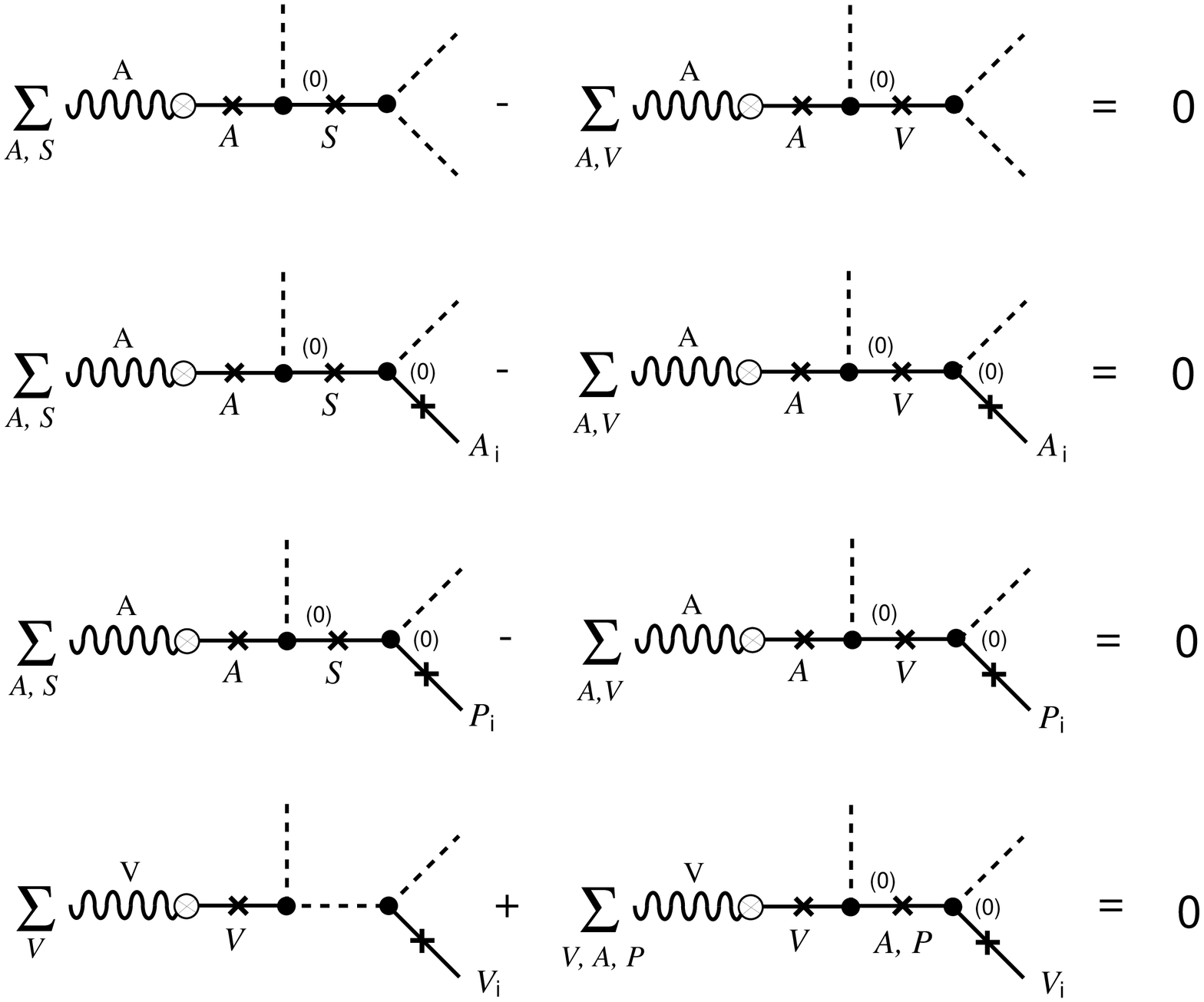}
\epsfxsize 4in
\centerline{\epsfbox{newMfourcurrentsb.eps}}
\caption{Sum rules for four-point functions with one external current
(\Eq{oneextsconcomp2}). Dotted lines are pions. Crosses represent
mass-squared insertions.}
\endcaption
\endfigure

\vskip0.2in

\noindent{\underbar {\it Four-point functions (one external current):}}

\offparens
$$\EQNalign{ 
&{\sum_{\sss S,A}}{G_{{\sss S}\pi}}{G^{\sss ({0})}_{\sss S{A}}}{F_{\sss A}}
({M_{\sss A}^{2}}-2{M_{\sss S}^{2}})
-{\sum_{\sss V,A}}{G_{{\sss V}\pi}}{G^{\sss ({0})}_{\sss V{A}}}{F_{\sss A}}
({M_{\sss A}^{2}}-2{M_{\sss V}^{2}})=0
\EQN oneextsconcomp2;a\cr
&{\sum_{\sss S,A}}{G^{\sss ({0})}_{\sss S{A_i}}}{G^{\sss ({0})}_{\sss S{A}}}
{F_{\sss A}}({M_{\sss A_i}^{2}}+{M_{\sss A}^{2}}-2{M_{\sss S}^{2}})-
{\sum_{\sss V,A}}{G^{\sss ({0})}_{\sss V{A_i}}}{G^{\sss ({0})}_{\sss V{A}}}
{F_{\sss A}}({M_{\sss A_i}^{2}}+{M_{\sss A}^{2}}-2{M_{\sss V}^{2}})=0
\EQN oneextsconcomp2;b\cr
&{\sum_{\sss S,A}}{G^{\sss ({0})}_{\sss S{P_i}}}{G^{\sss ({0})}_{\sss S{A}}}
{F_{\sss A}}({M_{\sss P_i}^{2}}+{M_{\sss A}^{2}}-2{M_{\sss S}^{2}})-
{\sum_{\sss V,A}}{G^{\sss ({0})}_{\sss V{P_i}}}{G^{\sss ({0})}_{\sss V{A}}}
{F_{\sss A}}({M_{\sss P_i}^{2}}+{M_{\sss A}^{2}}-2{M_{\sss V}^{2}})=0
\EQN oneextsconcomp2;c\cr
&{\sum_{\sss V}}{G_{{\sss {V_i}}\pi}}{G_{{\sss {V}}\pi}}{F_{\sss V}}
({M_{\sss V_i}^{2}}+{M_{\sss V}^{2}})+
{\sum_{\sss A,V}}{G^{\sss ({0})}_{\sss {V_i}A}}{G^{\sss ({0})}_{\sss {V}A}}
{F_{\sss V}}({M_{\sss V_i}^{2}}+{M_{\sss V}^{2}}-2{M_{\sss A}^{2}})\cr
&+{\sum_{\sss P,V}}{G^{\sss ({0})}_{\sss {V_i}P}}{G^{\sss ({0})}_{\sss {V}P}}
{F_{\sss V}}({M_{\sss V_i}^{2}}+{M_{\sss V}^{2}}-2{M_{\sss P}^{2}})=0
\EQN oneextsconcomp2;d\cr}
$$

\vskip0.2in

\noindent{\underbar {\it Four-point functions (two external currents):}}

\offparens
$$\EQNalign{ 
&{\sum_{\sss A,S,A'}}{F_{\sss A}}
{G^{\sss ({0})}_{\sss S{A}}}{G^{\sss ({0})}_{\sss S{A'}}}
{F_{\sss A'}}({M_{\sss A}^{2}}+{M_{\sss A'}^{2}}-2{M_{\sss S}^{2}})\cr
&-{\sum_{\sss A,V,A'}}{F_{\sss A}}
{G^{\sss ({0})}_{\sss V{A}}}{G^{\sss ({0})}_{\sss V{A'}}}
{F_{\sss A'}}({M_{\sss A}^{2}}+{M_{\sss A'}^{2}}-2{M_{\sss V}^{2}})=0
\EQN twoextsconcomp2;a\cr
&{\sum_{\sss V,V'}}{F_{\sss V}}
{G_{{\sss {V}}\pi}}{G_{{\sss {V'}}\pi}}{F_{\sss V'}}
({M_{\sss V}^{2}}+{M_{\sss V'}^{2}})+
{\sum_{\sss V,A,V'}}{F_{\sss V}}
{G^{\sss ({0})}_{\sss {V}A}}{G^{\sss ({0})}_{\sss {V'}A}}
{F_{\sss V'}}({M_{\sss V}^{2}}+{M_{\sss V'}^{2}}-2{M_{\sss A}^{2}})\cr
&+{\sum_{\sss V,P,V'}}{F_{\sss V}}
{G^{\sss ({0})}_{\sss {V}P}}{G^{\sss ({0})}_{\sss {V'}P}}
{F_{\sss V'}}({M_{\sss V}^{2}}+{M_{\sss V'}^{2}}-2{M_{\sss P}^{2}})=0.
\EQN twoextsconcomp2;b\cr}
$$ 
Notice that the first sum rule is the second spectral function sum
rule, {\it SFSR2}. Here again we see that there are an infinite number
of additional sum rules which are at precisely the same level of
theoretical rigor.  The new sum rules are illustrated diagramatically
in \Fig{yetagain.eps} (\Eq{sconcomp2}),
\Fig{newMfourcurrentsb.eps} (\Eq{oneextsconcomp2}) and
\Fig{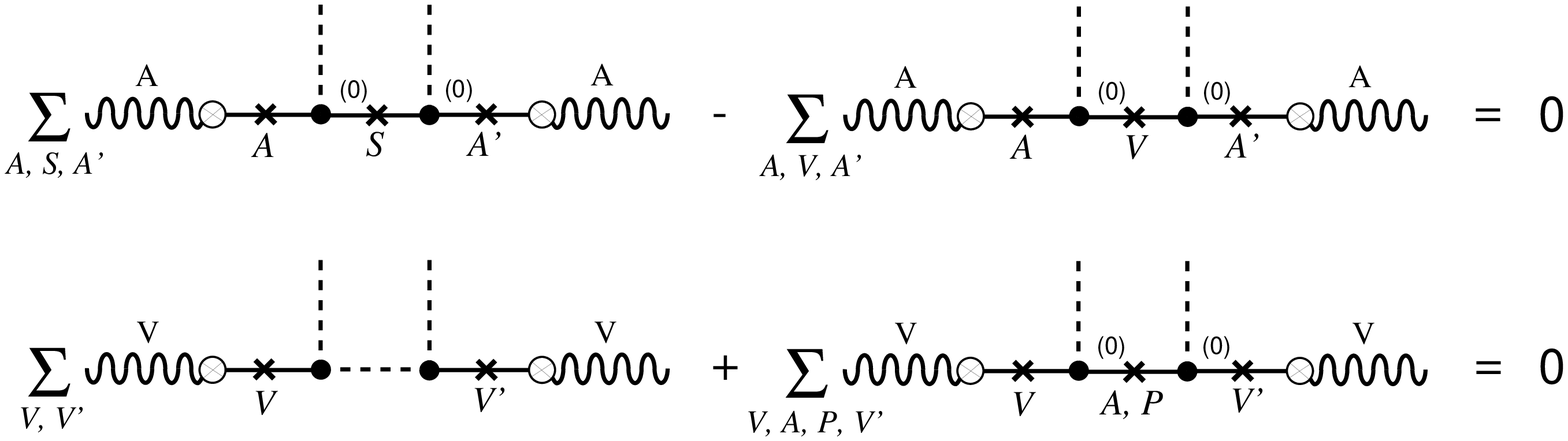} (\Eq{twoextsconcomp2}). (Many additional
sum rules can be constructed by contracting the meson masses, decay
constants and axial couplings.)

\figure{newMfourcurrentsc.eps}
\epsfxsize 5in
\centerline{\epsfbox{newMfourcurrentsc.eps}}
\caption{Sum rules for four-point functions with two external currents
(\Eq{twoextsconcomp2}). Dotted lines are pions. Crosses represent
mass-squared insertions.}
\endcaption
\endfigure

\autoparens
\vskip0.1in
\noindent {\twelvepoint{\bf 4.\quad Large-$\nc$ Asymptotics}}
\vskip0.1in

Of the chiral sum rules derived, there are only five which involve
decay constants and axial couplings with only $\lambda =0$
components. They are:

\offparens
$$\EQNalign{ 
&{\sum _{\sss V}}{F_{\sss V}^2}-{\sum _{\sss A}}{F_{\sss A}^2}=\yo1
\EQN summrelfir;a\cr
&{\sum_{\sss V}}{F_{\sss V}^2}{M_{\sss V}^{2}}-
{\sum_{\sss A}}{F_{\sss A}^2}{M_{\sss A}^{2}}=0
\EQN summrelfir;b\cr
&{\sum_{\sss V}}{F_{\sss V}}{G_{{\sss V}\pi}}=\yo1
\EQN summrelfir;c\cr
&{\sum_{\sss S}}{G^2_{{\sss S}\pi}}+{\sum_{\sss V}}{G^2_{{\sss V}\pi}}=\yo1
\EQN summrelfir;d\cr
&{\sum_{\sss S}}{G_{{\sss S}\pi}^2}{M_{\sss S}^2}-
{\sum_{\sss V}}{G_{{\sss V}\pi}^2}{M_{\sss V}^2}=0.
\EQN summrelfir;e\cr}
$$
\autoparens 
We have seen that the SFSR's, \Eq{summrelfir;a} and \Eq{summrelfir;b},
determine the asymptotic behavior of the two-point function
${\Pi_{\sss LR}}$ and in turn allow the construction of the full
correlation function in the large-$\nc$ limit. In this section, the
correlators relevant to the other three sum rules will be identified
and constructed using dispersion relations in precisely the same
manner.

\vskip0.1in
\noindent {\twelvepoint{\it 4.1\quad The Pion Vector Form Factor}}

Consider first a familiar product of three currents.  The
vector form factor of the pion is defined by

\offparens
$$
\bra{{\pi^a}p'}{{\rm V}^b_\mu}\ket{{\pi^c}p}={\epsilon^{abc}}
({p'_\mu}+{p_\mu}){F_{\sss V}}(q^2).
\EQN pionffdefined$$
\autoparens 
Charge conservation implies ${{F_{\sss V}}(0)}=1$.  The pion form
factor is an analytic function in the complex-t plane with a cut
extending along the real axis from the origin (the $t$-channel
$\pi\pi$ threshold in the chiral limit). Therefore, ${{F_{\sss
V}}(t)}$ satisfies a dispersion relation. Say the asymptotic behavior
of ${{F_{\sss V}}(t)}$ is given by:

\offparens
$$
{{F_{\sss V}}(t)}\mapright{t\rightarrow\infty}\, {t^{\bf m}},
\EQN asymbeh$$
\autoparens 
where the integer ${\bf m}$ determines the number of subtractions.  It
is possible to show in QCD that ${\bf m}\leq 0$\ref{ecker}. Hence
there is at most one subtraction.  With one subtraction at the origin
and making use of ${{F_{\sss V}}(0)}=1$ we have:

\offparens
$$
{{F_{\sss V}}(t)}=1+{t\over{\pi}}\int{{Im\; {{F_{\sss V}}({t'})}d{t'}}\over{
{t'}({t'}-t-i\epsilon)}}.
\EQN pionff2ones$$
\autoparens 
In the large-$\nc$ limit, the form factor is determined by an infinite
number of meson exchanges between the vector current and the $\pi \pi$
channel. Therefore, the absorptive part may be written as the formal
sum over states:

\offparens
$$
Im\; {{F_{\sss V}}(t)}=
{\sum_{\sss V}}{{{F_{\sss V}}{G_{{\sss V}\pi}}t\pi}\over\yo1}
\delta (t-{M_{\sss V}^2}) 
\EQN pionff3$$
\autoparens 
which gives

$$
{{F_{\sss V}}(t)}=1 +
{\sum_{\sss V}}{{{F_{\sss V}}{G_{{\sss V}\pi}}}\over\yo1}
{t\over{({M_{\sss V}^2}-t)}}.
\EQN pionffexactconst$$
It is convenient to expand this form factor in inverse powers of $t$:

\offparens
$$
{{F_{\sss V}}(t)}\mapright{t\rightarrow\infty}\, {1\over\yo1}
\Bigl\{\yo1- {\sum_{\sss V}}{{{F_{\sss V}}{G_{{\sss V}\pi}}}}\Bigr\}
-{1\over\yo1}{\sum_{\sss n=1}^\infty}\Bigl\{{\sum_{\sss V}}
{{{F_{\sss V}}{G_{{\sss V}\pi}}}}{M_{\sss V}^{2n}}\Bigr\}{1\over{t^n}}.
\EQN asymbehfulfu$$
But we have derived the exact relation

$$
{\sum_{\sss V}}{F_{\sss V}}{G_{{\sss V}\pi}}=\yo1 .
\EQN pionff5$$
One can further verify directly from 
\Eq{sf1decay}, \Eq{picouple} and \Eq{physmasses;b} that
generally there is no inverse power, $n$, of $t$ for which

$$
{\sum_{\sss V}}{{{F_{\sss V}}{G_{{\sss V}\pi}}}}{M_{\sss V}^{2n}}
\EQN pionffscrels$$
vanishes. Hence ${\bf m}=-1$ and {\it the asymptotic behavior of the
pion form factor is determined by chiral symmetry in the large-$\nc$
limit}. The pion form factor satisfies an unsubtracted dispersion
relation and is given by:

$$
{{F_{\sss V}}(t)}=
{\sum_{\sss V}}{{{F_{\sss V}}{G_{{\sss V}\pi}}}\over\yo1}
{{M_{\sss V}^2}\over{({M_{\sss V}^2}-t)}},
\EQN pionffexactubsub$$
subject to the normalization condition ${{F_{\sss V}}(0)}=1$.  We
emphasize that this is the {\it exact} form factor in the large-$\nc$
limit$^4$\vfootnote4{This form factor was studied in the context of
the large-$\nc$ approximation in \Ref{pich}. There the asymptotic
constraint, \Eq{pionff5}, was assumed rather than derived.}.

\vskip0.1in
\noindent {\twelvepoint{\it 4.2\quad Pion-pion Scattering}}

Next we consider correlation functions which are the time-ordered
product of four currents. We will focus on the forward $\pi-\pi$
scattering amplitudes in a basis of definite t-channel isospin as a
function of the crossing-odd variable $\nu=s-u$. The crossing-odd
amplitude is pure $I_t=1$ and the crossing-even amplitude contains
$I_t=0$ and $I_t=2$.  The analytic structure of the $\pi-\pi$
scattering amplitudes is well known\ref{pipi}. Leading order in chiral
perturbation theory\ref{gass} determines the small-$\nu$ behavior at
$t=0$:

\offparens 
$$\EQNalign{ 
& {{T^{\sss 1}_{t}}(\nu,0)}={\nu\over\yo1} \EQN tchannamps;a\cr
&{{T^{\sss 0}_{t}}(\nu,0)}={{T^{\sss 2}_{t}}(\nu,0)}=0. \EQN tchannamps;b\cr}
$$\autoparens
Say the asymptotic behavior is given by

\offparens 
$$\EQNalign{ 
& {{T^{\sss 1}_{t}}(\nu,0)}\,\mapright{\nu\rightarrow\infty}\, 
{\nu^{2{\bf m_1}+1}}
\EQN asymptgeh;a\cr
&{{T^{\sss 2}_{t}}(\nu,0)}\,\mapright{\nu\rightarrow\infty}\, 
{\nu^{2{\bf m_2}}}
\EQN asymptgeh;b\cr
&{{T^{\sss 0}_{t}}(\nu,0)}\,\mapright{\nu\rightarrow\infty}\, 
{\nu^{2{\bf m_0}}}
\EQN asymptgeh;c\cr}
$$
\autoparens 
where the integers ${\bf m_i}$ determine the number of subractions.
The Froissart bound implies the upper bound ${\bf m_i}\leq
0$\ref{froi}.

We will consider first the $I_t=1$ amplitude, which is
crossing-odd. Saturating the Froissart bound (assuming the most severe
asymptotic behavior) we make one subtraction at ${\nu_0}=0$ which
yields

\offparens
$$
{{{T^{\sss 1}_{t}}(\nu,0)}\over \nu}=
\lim_{{\nu_0}\to 0}{{{T^{\sss 1}_{t}}({\nu_0},0)}\over {\nu_0}}
+{{2\nu^2}\over\pi}\int_0^\infty
{{d{\nu'}Im\;{T^{\sss 1}_{t}}(\nu',0)}\over {{\nu'^2}({\nu'^2}-{\nu^2})}}.
\EQN pionscattIonesub$$
\autoparens 
Using \Eq{tchannamps;a} and going to a basis of s-channel isospin
gives

\offparens
$$
{{{T^{\sss 1}_{t}}(\nu,0)}}=
{\nu\over\yo1}
+{{\nu^3}\over\pi}\int_0^\infty
{{{d{\nu'}}[{\ssty 
{2\over 3}}{Im\;{T^{\sss 0}_{s}}(\nu',0)}
  +{Im\;{T^{\sss 1}_{s}}(\nu',0)}
-{\ssty {5\over 3}}{Im\;{T^{\sss 2}_{s}}(\nu',0)}]}
\over {{{\nu'^2}({\nu'^2}-{\nu^2})}}}.
\EQN pionscattIonesub2$$
In the large-$\nc$ limit we can write the absorptive
parts as formal sums over single-particle states:

$$\EQNalign{ 
&{Im\;{T^{\sss 1}_{s}}(\nu,0)}
={\sum_{\sss V}}
{{4\pi{M_{\sss V}^4}{G_{{\sss V}\pi}^2}}\over{{F_\pi^4}}}
\delta ({\nu} -{2M_{\sss V}^2})\EQN BW;a\cr
&{Im\;{T^{\sss 0}_{s}}(\nu,0)}
={\sum_{\sss S}}
{{6\pi{M_{\sss S}^4}{G_{{\sss S}\pi}^2}}\over{{F_\pi^4}}}
\delta ({\nu} -{2M_{\sss S}^2}).\EQN BW;b\cr}
$$\autoparens 
Of course ${Im\;{T^{\sss 2}_{s}}}$ vanishes since there are no $I=2$
mesons in the large-$\nc$ limit\ref{witten}. It immediately follows that

\offparens
$$
{{{T^{\sss 1}_{t}}(\nu,0)}}=
{\nu\over\yo1}
+{{\nu^3}\over{F_\pi^4}}\Bigl\lbrack
{\sum_{\sss V}}{{G_{{\sss V}\pi}^2}\over{{4M_{\sss V}^4}-{\nu^2}}}+  
{\sum_{\sss S}}{{G_{{\sss S}\pi}^2}\over{{4M_{\sss S}^4}-{\nu^2}}}  
\Bigr\rbrack .
\EQN pionscattIonesub3$$
Expanding in inverse powers of $\nu$ gives:

$$\eqalign{ 
{{{T^{\sss 1}_{t}}(\nu,0)}}\mapright{\nu\rightarrow\infty}\, 
&{1\over{F_\pi^4}}\Bigl\{\yo1 -
{\sum_{\sss V}}{{G_{{\sss V}\pi}^2}}-
{\sum_{\sss S}}{{G_{{\sss S}\pi}^2}}\Bigr\}{\nu}\cr
-&{1\over{F_\pi^4}}{\sum_{\sss n=1}^\infty}\Bigl\{
{\sum_{\sss V}}{{G_{{\sss V}\pi}^2}}{(4M_{\sss V}^4)^n}+
{\sum_{\sss S}}{{G_{{\sss S}\pi}^2}}{(4M_{\sss S}^4)^n}
\Bigr\}{1\over {\nu^{2n-1}}}.\cr}
\EQN pionscattIonesubexp$$
But we have derived the exact relation

\offparens
$$
{\sum_{\sss V}}{G_{{\sss V}\pi}^2}+{\sum_{\sss S}}{G_{{\sss S}\pi}^2}=
{F_\pi^2}. \EQN avsrpf$$
\autoparens 
Moreover, the coefficients of all inverse powers of $\nu$ are positive
definite.  Hence it is clear that ${\bf m_1}=-1$.  Therefore, chiral
symmetry determines that ${T^{\sss 1}_{t}}(\nu,0)$ satisfies
an unsubtracted dispersion relation which yields

\offparens
$$
{{{T^{\sss 1}_{t}}(\nu,0)}}=
{{4\nu}\over{F_\pi^4}}\Bigl\lbrack
{\sum_{\sss V}}
{{{G_{{\sss V}\pi}^2}{M_{\sss V}^4}}\over{{4M_{\sss V}^4}-{\nu^2}}}+  
{\sum_{\sss S}}
{{{G_{{\sss S}\pi}^2}{M_{\sss S}^4}}\over{{4M_{\sss S}^4}-{\nu^2}}}  
\Bigr\rbrack 
\EQN pionscattIoneunsub$$
subject to the normalization condition $\lim_{{\nu_0}\to 0}{{{T^{\sss
1}_{t}}({\nu_0},0)}/{\nu_0}}={1/\yo1}$.  This condition (\Eq{avsrpf})
is sometimes expressed via the optical theorem as

$$
{1\over\yo1}=
{1\over{\pi}}\int{{d{\nu}}\over{\nu}} 
[{\ssty {1\over 3}}{\sigma_{s}^{\sss 0}}({\nu} )
  +{\ssty {1\over 2}}{\sigma_{s}^{\sss 1}}({\nu} )
-{\ssty {5\over 6}}{\sigma_{s}^{\sss 2}}({\nu} ) ]
\EQN awdisp
$$\autoparens which is the Adler-Weisberger sum rule for $\pi-\pi$
scattering$^5$\vfootnote5{In effect, the sum rules of \Eq{aw1comp2}
comprise the complete set of Adler-Weisberger sum rules for 
$\pi$-meson scattering.}.  Next we turn to the
$I_t=2$ amplitude. The $I_t=0$ and $I_t=2$ amplitudes are 
crossing even. With one subtraction at ${\nu_0}=0$ we have

\offparens
$$
{{{T^{\sss 2}_{t}}(\nu,0)}}=
\lim_{{\nu_0}\to 0}{{{T^{\sss 2}_{t}}({\nu_0},0)}}
+{{2\nu^2}\over\pi}\int_0^\infty
{{d{\nu'}Im\;{T^{\sss 2}_{t}}(\nu',0)}\over {{\nu'}({\nu'^2}-{\nu^2})}}.
\EQN pionscattItwosub$$
Using \Eq{tchannamps;b} yields

\offparens
$$
{{{T^{\sss 2}_{t}}(\nu,0)}}=
{{\nu^2}\over\pi}\int_0^\infty
{{{d{\nu'}}[{\ssty 
{2\over 3}}{Im\;{T^{\sss 0}_{s}}(\nu',0)}
  -{Im\;{T^{\sss 1}_{s}}(\nu',0)}
+{\ssty {1\over 3}}{Im\;{T^{\sss 2}_{s}}(\nu',0)}]}
\over {{{\nu'}({\nu'^2}-{\nu^2})}}}
\EQN pionscattItwosub2$$
in a basis of s-channel isospin. Saturating with the sum over states
gives

\offparens
$$
{{{T^{\sss 2}_{t}}(\nu,0)}}=
-{{2{\nu^2}}\over{F_\pi^4}}\Bigl\lbrack
{\sum_{\sss V}}
{{{G_{{\sss V}\pi}^2}{M_{\sss V}^2}}\over{{4M_{\sss V}^4}-{\nu^2}}}- 
{\sum_{\sss S}}
{{{G_{{\sss S}\pi}^2}{M_{\sss S}^2}}\over{{4M_{\sss S}^4}-{\nu^2}}}  
\Bigr\rbrack .
\EQN pionscattItwounsub$$
Expanding in inverse powers of $\nu$ yields:

$$\eqalign{ 
{{{T^{\sss 2}_{t}}(\nu,0)}}\mapright{\nu\rightarrow\infty}\, 
&{2\over{F_\pi^4}}\Bigl\{
{\sum_{\sss V}}{{G_{{\sss V}\pi}^2}}{M_{\sss V}^2}-
{\sum_{\sss S}}{{G_{{\sss S}\pi}^2}}{M_{\sss S}^2}\Bigr\}\cr
+&{1\over{F_\pi^4}}{\sum_{\sss n=1}^\infty}\Bigl\{
{\sum_{\sss V}}{{G_{{\sss V}\pi}^2}}{({2M_{\sss V}^2})^{2n+1}}-
{\sum_{\sss S}}{{G_{{\sss S}\pi}^2}}{({2M_{\sss S}^2})^{2n+1}}
\Bigr\}{1\over {\nu^{2n}}}.\cr}
\EQN pionscattItwosubexp$$
But we have derived the exact relation

\offparens
$$
{\sum_{\sss V}}{G_{{\sss V}\pi}^2}{M_{\sss V}^2}=
{\sum_{\sss S}}{G_{{\sss S}\pi}^2}{M_{\sss S}^2}.
\EQN scpasd$$
\autoparens 
One can further verify directly from 
\Eq{picouple} and \Eq{physmasses} that
generally there is no power $n$ for which

\offparens
$$
{\sum_{\sss V}}{G_{{\sss V}\pi}^2}{M_{\sss V}^{4n+2}}=
{\sum_{\sss S}}{G_{{\sss S}\pi}^2}{M_{\sss S}^{4n+2}}.
\EQN scpasdasdf$$
\autoparens 
It is clear that ${\bf m_2}=-1$ and chiral symmetry determines that
${T^{\sss 2}_{t}}(\nu,0)$ satisfies an unsubtracted
dispersion relation which yields

\offparens
$$
{{{T^{\sss 2}_{t}}(\nu,0)}}=
-{{8}\over{F_\pi^4}}\Bigl\lbrack
{\sum_{\sss V}}
{{{G_{{\sss V}\pi}^2}{M_{\sss V}^6}}\over{{4M_{\sss V}^4}-{\nu^2}}}-  
{\sum_{\sss S}}
{{{G_{{\sss S}\pi}^2}{M_{\sss S}^6}}\over{{4M_{\sss S}^4}-{\nu^2}}}  
\Bigr\rbrack 
\EQN pionscattItwounsubwa$$
subject to the normalization condition $\lim_{{\nu_0}\to 0} {{T^{\sss
2}_{t}}({\nu_0},0)}=0$.  This condition (\Eq{scpasd}) is sometimes
expressed via the optical theorem as the superconvergent sum rule:

$$
0=\int{{d{\nu}}} 
[{\ssty {1\over 3}}{\sigma_{s}^{\sss 0}}({\nu} )
-{\ssty {1\over 2}}{\sigma_{s}^{\sss 1}}({\nu} )
+{\ssty {1\over 6}}{\sigma_{s}^{\sss 2}}({\nu} ) ].
\EQN scdisp
$$\autoparens 

Next we turn to the $I_t=0$ amplitude. With one subtraction at
${\nu_0}=0$ we have

\offparens
$$
{{{T^{\sss 0}_{t}}(\nu,0)}}=
\lim_{{\nu_0}\to 0}{{{T^{\sss 0}_{t}}({\nu_0},0)}}
+{{2\nu^2}\over\pi}\int_0^\infty
{{d{\nu'}Im\;{T^{\sss 0}_{t}}(\nu',0)}\over {{\nu'}({\nu'^2}-{\nu^2})}}.
\EQN pionscattIzerosub$$
Again using \Eq{tchannamps;b} gives

\offparens
$$
{{{T^{\sss 0}_{t}}(\nu,0)}}=
{{\nu^2}\over\pi}\int_0^\infty
{{{d{\nu'}}[{\ssty {2\over 3}}{Im\;{T^{\sss 0}_{s}}(\nu',0)}
  +2{Im\;{T^{\sss 1}_{s}}(\nu',0)}
+{\ssty {10\over 3}}{Im\;{T^{\sss 2}_{s}}(\nu',0)}]}
\over {{{\nu'}({\nu'^2}-{\nu^2})}}}
\EQN pionscattIzerosub2$$
in a basis of s-channel isospin.
Saturating with the sum over states gives

\offparens
$$
{{{T^{\sss 0}_{t}}(\nu,0)}}=
{{2\nu^2}\over{F_\pi^4}}\Bigl\lbrack
2{\sum_{\sss V}}
{{{G_{{\sss V}\pi}^2}{M_{\sss V}^2}}\over{{4M_{\sss V}^4}-{\nu^2}}}+ 
{\sum_{\sss S}}
{{{G_{{\sss S}\pi}^2}{M_{\sss S}^2}}\over{{4M_{\sss S}^4}-{\nu^2}}}  
\Bigr\rbrack .
\EQN pionscattIzerounsub$$
Expanding in inverse powers of $\nu$ yields:

$$\eqalign{ {{{T^{\sss 0}_{t}}(\nu,0)}}\mapright{\nu\rightarrow\infty}\,
-&{2\over{F_\pi^4}}\Bigl\{ 2{\sum_{\sss V}}{{G_{{\sss
V}\pi}^2}}{M_{\sss V}^2}+ {\sum_{\sss S}}{{G_{{\sss S}\pi}^2}}{M_{\sss
S}^2}\Bigr\}\cr -&{1\over{F_\pi^4}}{\sum_{\sss n=1}^\infty}\Bigl\{
2{\sum_{\sss V}}{{G_{{\sss V}\pi}^2}}{({2M_{\sss V}^2})^{2n+1}}+
{\sum_{\sss S}}{{G_{{\sss S}\pi}^2}}{({2M_{\sss S}^2})^{2n+1}}
\Bigr\}{1\over {\nu^{2n}}}.\cr} 
\EQN pionscattIzerosubexp$$ 
Clearly chiral symmetry does not constrain the asymptotic behavior and
so ${\bf m_0}=0$ (maximal strength consistent with the Froissart
bound). However, chiral symmetry gives an interesting interpretation
to the leading coefficient of the $I_t=0$ amplitude in the $1/\nu$
expansion.  Using \Eq{pomestuff}, \Eq{scpasd} and 
\Eq{pionscattIzerosubexp} gives the remarkable equation

\offparens
$$ {{{T^{\sss 0}_{t}}(\nu,0)}}\mapright{\nu\rightarrow\infty}\,
{1\over{F_\pi^2}}\bra{{\pi}_a} {{\hat M}^2_{\sss
\bf{(2,2)}}}\ket{{\pi}_b}\;{\delta_{ab}}\; +\; O({{\nu^{-2}}}).
\EQN pomestuffpipi
$$ 
Since the $I_t=1$ and $I_t=2$ amplitudes fall off at large $\nu$, it
follows that the asymptotic behavior of the total cross-section is
determined by the symmetry breaking part of the mass-squared
matrix$^6$\vfootnote6{Attempts to construct ${{\hat M}^2_{\sss
\bf{(2,2)}}}$ explicitly in QCD are discussed in the fascinating
papers of \Ref{suss} (see also \Ref{beanefirst}).}.  This result was
known long ago\ref{algebraic}\ref{suss}. Here it is seen to be exact
in the large-$\nc$ limit.

\vskip0.1in \noindent {\twelvepoint{\it 4.3\quad Summary}}

We have used the relations,

\offparens
$$\EQNalign{ 
&{\sum_{\sss V}}{F_{\sss V}}{G_{{\sss V}\pi}}=\yo1
\EQN summrel;a\cr
&{\sum_{\sss S}}{G^2_{{\sss S}\pi}}+{\sum_{\sss V}}{G^2_{{\sss V}\pi}}=\yo1
\EQN summrel;b\cr
&{\sum_{\sss V}}{G_{{\sss V}\pi}^2}{M_{\sss V}^2}=
{\sum_{\sss S}}{G_{{\sss S}\pi}^2}{M_{\sss S}^2}
\EQN summrel;c\cr}
$$
\autoparens 
which were derived from the pion chiral representation (and the chiral
algebra in an appendix) to determine the asymptotic behavior
of the pion vector form factor and the forward $\pi-\pi$ scattering
amplitudes, respectively:

\offparens 
$$\EQNalign{ 
&{{F_{\sss V}}(t)}\mapright{t\rightarrow\infty}\, {t^{-1}}
\EQN summasymptbeh;a\cr
& {{T^{\sss 1}_{t}}(\nu,0)}\,\mapright{\nu\rightarrow\infty}\, {\nu^{-1}}
\EQN summasymptbeh;b\cr
&{{T^{\sss 2}_{t}}(\nu,0)}\,\mapright{\nu\rightarrow\infty}\, {\nu^{-2}}
\EQN summasymptgeh;c\cr
& {{T^{\sss 0}_{t}}(\nu,0)}\,\mapright{\nu\rightarrow\infty}\, {\nu^{0}}.
\EQN summasymptbeh;d\cr}
$$
\autoparens 
Using dispersion theory the precise large-$\nc$ form factor and
forward scattering amplitudes were constructed:

\offparens 
$$\EQNalign{ 
&{{F_{\sss V}}(t)}=
{\sum_{\sss V}}{{{F_{\sss V}}{G_{{\sss V}\pi}}}\over\yo1}
{{M_{\sss V}^2}\over{({M_{\sss V}^2}-t)}};\qquad
{{F_{\sss V}}(0)}=1
\EQN summamps;a\cr
&{{{T^{\sss 1}_{t}}(\nu,0)}}=
{{4\nu}\over{F_\pi^4}}\Bigl\lbrack
{\sum_{\sss V}}
{{{G_{{\sss V}\pi}^2}{M_{\sss V}^4}}\over{{4M_{\sss V}^4}-{\nu^2}}}+  
{\sum_{\sss S}}
{{{G_{{\sss S}\pi}^2}{M_{\sss S}^4}}\over{{4M_{\sss S}^4}-{\nu^2}}}  
\Bigr\rbrack ; \qquad
\lim_{{\nu_0}\to 0}{{{T^{\sss 1}_{t}}({\nu_0},0)}\over{\nu_0}}={1\over\yo1}
\EQN summamps;b\cr
&{{{T^{\sss 2}_{t}}(\nu,0)}}=
-{{8}\over{F_\pi^4}}\Bigl\lbrack
{\sum_{\sss V}}
{{{G_{{\sss V}\pi}^2}{M_{\sss V}^6}}\over{{4M_{\sss V}^4}-{\nu^2}}}-  
{\sum_{\sss S}}
{{{G_{{\sss S}\pi}^2}{M_{\sss S}^6}}\over{{4M_{\sss S}^4}-{\nu^2}}}  
\Bigr\rbrack ;\qquad
\lim_{{\nu_0}\to 0} {{T^{\sss 2}_{t}}({\nu_0},0)}=0.
\EQN summamps;c\cr
&{{{T^{\sss 0}_{t}}(\nu,0)}}=
{{2\nu^2}\over{F_\pi^4}}\Bigl\lbrack
2{\sum_{\sss V}}
{{{G_{{\sss V}\pi}^2}{M_{\sss V}^2}}\over{{4M_{\sss V}^4}-{\nu^2}}}+ 
{\sum_{\sss S}}
{{{G_{{\sss S}\pi}^2}{M_{\sss S}^2}}\over{{4M_{\sss S}^4}-{\nu^2}}}  
\Bigr\rbrack \EQN summamps;d\cr}
$$
\autoparens 

In similar fashion one can derive the asymptotic behavior of and
construct other form factors and forward scattering amplitudes from
the remaining sum rules.  In order to make contact between the results
of this paper and phenomenology, the infinite space of meson states
must be truncated to a few low-lying states. The constraints implied
by finite dimensional saturation schemes on the low-energy constants
of chiral perturbation theory are discussed elsewhere\ref{beane}. A
phenomenological study of the form factors and forward scattering
amplitudes in turn requires going beyond the zero-width form of the
large-$\nc$ correlators which necessitates subsuming higher order
effects in the 1/$\nc$ expansion. Considerations similar to those made
here lead to sum rules involving other nonconserved currents (e.g.
scalar and pseudoscalar currents). Chiral symmetry constraints at
infinite momentum for the ground state large-$\nc$ baryons have been
studied in \Ref{strong} and \Ref{beane2}.  Analogs of the SFSR's for
heavy mesons have been studied in the Large-$\nc$ limit in \Ref{luty}.

\vskip0.1in
\noindent {\twelvepoint{\bf 5.\quad Conclusion}}
\vskip0.1in

The remarkable fact that operator product expansion coefficients do
not feel vacuum properties implies that the full global symmetry of an
underlying theory can have profound algebraic relevance in the
low-energy theory even if the full global symmetry is spontaneously
broken\ref{ope}. The spectral function sum rules are classic examples
of this fact in QCD. Traditionally these sum rules are derived using
the OPE.

In this paper I rederived the SFSR's using chiral symmetry directly,
both from the representation theory and from the chiral algebra (see
appendix). The advantage of the symmetry method is that all
consequences of chiral symmetry can be derived at once for arbitrary
$n$-point functions.  Hence using this method I derived an infinite
number of new sum rules for $3$- and $4$-point functions in the
large-$\nc$ limit.  These sum rules are at precisely the same level of
rigor as the SFSR's.  The sum rules constrain the asymptotic behavior
of correlation functions and thereby allow the derivation of exact
large-$\nc$ correlators using dispersion relations. It is somewhat
surprising that there are fundamental constraints on the asymptotic
behavior of scattering amplitudes which go beyond what is implied by
unitarity (via the Froissart bound).  Of particular interest is the
interplay between the asymptotic behavior of the total cross-section
and the symmetry breaking part of the low-energy Hamiltonian or
mass-squared matrix.

Generally the order parameter of chiral symmetry breaking in
large-$\nc$ QCD transforms as a sum of $\bf{(2,2)}$ and $\bf{(3,3)}$
representations. However, I found that the $\bf{(3,3)}$ part of the
low-energy Hamiltonian is proportional to an OPE coefficient which
vanishes in QCD. This constitutes a proof that the order parameter of
chiral symmetry breaking in large-$\nc$ QCD transforms purely as
$\bf{(2,2)}$, a result which is consistent with what is assumed in the
Coleman-Witten proof of chiral symmetry breaking at
large-$\nc$\ref{cole}.

The results of this paper demonstrate the fundamental importance of the
infinite momentum frame in exploiting chiral symmetry constraints in
QCD. Given the deep Euclidean nature of the OPE it is not surprising
that the infinite momentum frame is relevant in extracting
consequences of the full chiral group in the low-energy theory. It
would be interesting to verify some of the new sum rules for $3$- and
$4$-point functions using the OPE method of
\Ref{ope}.

\vskip0.1in
\noindent {\twelvepoint{\bf Acknowledgements}}
\vskip0.1in

\noindent This work was supported by the U.S. Department of Energy grant
DE-FG02-93ER-40762 (at Maryland) and DE-FG03-97ER-41014 (at
Washington).  I thank Markus Luty for valuable conversations.

\vskip0.1in
\noindent {\twelvepoint{\bf Appendix}}
\vskip0.1in

In this appendix I give an algebraic derivation of the basic
sum rules, \Eq{newmainaw1} -\Eq{mainaw1as} and \Eq{genmasssumsol}.
Consider the commutators involving the chiral charges and
the vector and axialvector currents:

\offparens
$$\EQNalign{ 
&[{Q_{5}^{a}},{{\rm V}_{\mu}^b}]=i{\epsilon^{abc}}{{\rm A}_{\mu}^c}
\EQN aandvagain;a\cr
&[{Q_{5}^{a}},{{\rm A}_{\mu}^b}]=i{\epsilon^{abc}}{{\rm V}_{\mu}^c}
\EQN aandvagain;b\cr
&[{Q^{5}_{a}},{Q^{5}_{b}}]=i{\epsilon_{abc}}{T_c}.
\EQN aandvagain;c\cr}
$$\autoparens We can take matrix elements of these commutators between
physical states at infinite momentum.  Consider, for instance,

\offparens
$$
\bra{0}[{Q_{5}^{a}},{{\rm V}_{\mu}^b}]{\ket{\Pi^e}_{\sss i}}
=i{\epsilon^{abc}}
\bra{0}{{\rm A}_{\mu}^c}{\ket{\Pi^e}_{\sss i}}.
\EQN currentalg1A
$$\autoparens 
Using ${Q_{5}^{a}}\ket{0}=0$ and inserting a complete set of states gives

\offparens
$$
-{\sum_{\sss j=1}^{\sss m}}\bra{0}{{\rm V}_{\mu}^b}{\ket{V^f}_{\sss j}}
{_{\sss\,\,\, j}}{\bra{V^f}}
{Q_{5}^{a}}{\ket{\Pi^e}_{\sss i}}
=i{\epsilon^{abc}}
\bra{0}{{\rm A}_{\mu}^c}{\ket{\Pi^e}_{\sss i}}.
\EQN currentalg1B
$$\autoparens Here we have also used helicity conservation. 
Using \Eq{aandvdefnew} and \Eq{coupgenrel} we then have

\offparens
$$
-{\sum_{\sss j=1}^{\sss m}}
({\delta^{bf}}{F_{\sss V_j}}{p_\mu})
(i{\epsilon^{aef}}{{G_{\sss {V_j}{\Pi_i}}}/{F_\pi}})
=i{\epsilon^{abc}}
({\delta^{ce}}{F_{\sss \Pi_i}}{p_\mu})
\EQN currentalg1C
$$\autoparens 
from which immediately follows

\offparens
$$
{\sum_{\sss j=1}^{\sss m}}{F_{\sss {V_j}}}{G_{\sss {V_j}{\Pi_i}}}=
{F_\pi}{F_{\sss \Pi_k}}.
\EQN currentalg1D
$$\autoparens 

Similarly, we have

\offparens
$$\EQNalign{ 
&\bra{0}[{Q_{5}^{a}},{{\rm A}_{\mu}^b}]{\ket{V^e}_{\sss i}}
=i{\epsilon^{abc}}\bra{0}{{\rm V}_{\mu}^c}{\ket{V^e}_{\sss i}}
\EQN aandvagainwh;a\cr
&\bra{0}[{Q_{5}^{a}},{{\rm A}_{\mu}^b}]{\ket{S}_{\sss i}}
=0
\EQN aandvagainwh;b\cr}
$$
which lead to

$$\EQNalign{ 
&{\sum_{\sss j=1}^{\sss m+n}}{F_{\sss {\Pi_j}}}{G_{\sss {V_i}{\Pi_j}}}=
{F_\pi}{F_{\sss {V_i}}} \EQN mainaw1;a\cr
&{\sum_{\sss j=1}^{\sss m+n}}{F_{\sss {\Pi_j}}}{G_{\sss {S_i}{\Pi_j}}}=0.
\EQN mainaw1;b\cr}
$$
Now contracting \Eq{currentalg1D} with ${F_{\sss \Pi_k}}$
and \Eq{mainaw1;a} with ${F_{\sss {V_i}}}$ implies

$$
{\sum_{\sss i=1}^{\sss m+n}}{F_{\sss {\Pi_i}}}{F_{\sss {\Pi_i}}}=
{\sum_{\sss i=1}^{\sss m}}{F_{\sss {V_i}}}{F_{\sss {V_i}}}.
\EQN newmainaw1again$$

Similarly we can take matrix elements of the commutator, \Eq{aandvagain;c}.
Consider, for instance, 

\offparens
$$
{_{\sss\,\,\, j}{\bra{V_e}}}
[{Q^{5}_{a}},{Q^{5}_{b}}]{\ket{V_d}_{\sss i}}
=i{\epsilon_{abc}}{_{\sss\,\,\, j}{\bra{V_e}}}{T_c}
{\ket{V_d}_{\sss i}}.
\EQN chiralalgB
$$\autoparens 
Inserting a complete set of states gives

\offparens
$$
{\sum_{\sss k=1}^{\sss m+n}}
{_{\sss\,\,\, j}{\bra{V_e}}}{Q^{5}_{a}}
{\ket{\Pi_l}_{\sss k}}
{_{\sss\,\,\, k}{\bra{\Pi_l}}}{Q^{5}_{b}}{\ket{V_d}_{\sss i}}-
({a\leftrightarrow b})
=i{\epsilon_{abc}}
{_{\sss\,\,\, j}{\bra{V_e}}}{T_{c}}
{\ket{V_d}_{\sss i}}
\EQN currentalg1asdf
$$\autoparens 
and using \Eq{gentensor} and \Eq{coupgenrel} we obtain

\offparens
$$
{\sum_{\sss k=1}^{\sss m+n}}
(i{\epsilon_{ale}}{{G_{\sss {V_j}{\Pi_k}}}/{F_\pi}})
(i{\epsilon_{lbd}}{{G_{\sss {V_i}{\Pi_k}}}/{F_\pi}})
-({a\leftrightarrow b})
=i{\epsilon_{abc}} (i{\delta_{ij}}{\epsilon_{ecd}}).
\EQN currentalg1asdfB
$$\autoparens 
Finally a bit of algebra gives

\offparens
$$
{\sum_{\sss i=1}^{\sss m+n}}{G_{\sss {V_j}{\Pi_i}}}{G_{\sss {V_k}{\Pi_i}}}=
\yo1{\delta_{jk}}.
\EQN mainaw1asyeta
$$
Similarly, we have

$$\EQNalign{ 
&{_{\sss\,\,\, j}{\bra{\Pi_e}}}
[{Q^{5}_{a}},{Q^{5}_{b}}]{\ket{\Pi_d}_{\sss i}}
=i{\epsilon_{abc}}{_{\sss\,\,\, j}{\bra{\Pi_e}}}{T_c}
{\ket{\Pi_d}_{\sss i}}
\EQN aandvagainwhgoof;a\cr
&{_{\sss\,\,\, j}{\bra{S}}}
[{Q^{5}_{a}},{Q^{5}_{b}}]{\ket{V_d}_{\sss i}}=0
\EQN aandvagainwhgoof;b\cr}
$$
which lead to

$$\EQNalign{ 
&{\sum_{\sss j=m+1}^{\sss m+n+o}}{G_{\sss {S_j}{\Pi_i}}}
{G_{\sss {S_j}{\Pi_k}}}+
{\sum_{\sss j=1}^{\sss m}}{G_{\sss {V_j}{\Pi_i}}}{G_{\sss {V_j}{\Pi_k}}}=
\yo1{\delta_{ik}} \EQN mainaw1asch;a\cr
&{\sum_{\sss i=1}^{\sss m+n}}{G_{\sss {S_j}{\Pi_i}}}{G_{\sss {V_k}{\Pi_i}}}=0.
\EQN mainaw1asch;b\cr}
$$
Hence we have reproduced \Eq{newmainaw1}-\Eq{mainaw1as} from \Eq{aandvagain}.

Taking ${{\hat M}^2_{\sss \bf{(2,2)}}}$ to transform as the zeroth
component of the $\bf{(2,2)}$ representation implies

\offparens
$$
[{Q^{5}_{a}}, {{\hat M}^2_{{\sss \bf{(2,2)}}o}}]=
i{{\hat M}^2_{{\sss \bf{(2,2)}}a}}
\EQN sbmmdefdaf1
$$
and

\offparens
$$
[{Q^{5}_{a}}, {{\hat M}^2_{{\sss \bf{(2,2)}}b}}]=
-i{\delta_{ab}}{{\hat M}^2_{{\sss \bf{(2,2)}}o}}.
\EQN sbmmdefdaf2
$$
Since
\offparens
$$
[{Q^{5}_{a}}, {{\hat M}^2_{{\sss \bf{(1,1)}}}}]=0
\EQN sbmmdefdaf2asdf
$$
we can write

\offparens
$$
[{Q^{5}_{a}},[{Q^{5}_{b}},{{\hat {M}}^2}]]=
{\delta_{ab}}{{\hat M}^2_{{\sss \bf{(2,2)}}}}.
\EQN sbmmdef
$$
Taking matrix elements of this commutation relation with the physical
states of the pion multiplet implies nontrivial sum rules. From

\offparens
$$\EQNalign{ 
&{_{\sss\,\,\, j}{\bra{\Pi_c}}}
[{Q^{5}_{a}},[{Q^{5}_{b}},{{\hat {M}}^2}]]
{\ket{\Pi_d}_{\sss i}}
\propto {\delta_{ab}}
\EQN genmasssum;a\cr
&{_{\sss\,\,\, j}{\bra{V_c}}}
[{Q^{5}_{a}},[{Q^{5}_{b}},{{\hat {M}}^2}]]
{\ket{V_d}_{\sss i}}
=0
\EQN genmasssum;b\cr
&{_{\sss\,\,\, j}{\bra{S}}}
[{Q^{5}_{a}},[{Q^{5}_{b}},{{\hat {M}}^2}]]
{\ket{S}_{\sss i}}
\propto {\delta_{ab}}
\EQN genmasssum;c\cr
&{_{\sss\,\,\, j}{\bra{S}}}
[{Q^{5}_{a}},[{Q^{5}_{b}},{{\hat {M}}^2}]]
{\ket{V_d}_{\sss i}}=0
\EQN genmasssum;d\cr}
$$ 
follow

\offparens
$$\EQNalign{ 
&{\sum_{\sss j=m+1}^{\sss m+n+o}}{G_{\sss {S_j}{\Pi_i}}}
{G_{\sss {S_j}{\Pi_k}}}
({M_{\sss \Pi_i}^{2}}+{M_{\sss \Pi_k}^{2}}-2{M_{\sss S_j}^{2}})\cr
&-{\sum_{\sss j=1}^{\sss m}}{G_{\sss {V_j}{\Pi_i}}}
{G_{\sss {V_j}{\Pi_k}}}
({M_{\sss \Pi_i}^{2}}+{M_{\sss \Pi_k}^{2}}-2{M_{\sss V_j}^{2}})=0
\EQN genmasssumsol2;a\cr
&{\sum_{\sss k=1}^{\sss m+n}}{G_{\sss {V_i}{\Pi_k}}}
{G_{\sss {V_j}{\Pi_k}}}
({M_{\sss V_i}^{2}}+{M_{\sss V_j}^{2}}-2{M_{\sss \Pi_k}^{2}})=0
\EQN genmasssumsol2;b\cr
&{\sum_{\sss k=1}^{\sss m+n}}{G_{\sss {S_i}{\Pi_k}}}
{G_{\sss {V_j}{\Pi_k}}}
({M_{\sss V_j}^{2}}-{M_{\sss S_i}^{2}})=0.
\EQN genmasssumsol2;c\cr}
$$ 
which are the sum rules of \Eq{genmasssumsol}.

\nosechead{References}
\ListReferences \vfill\supereject \end